\documentclass[12pt]{iopart}

\usepackage{iopams,amssymb}
\usepackage{graphicx}% Include figure files
\usepackage{psfrag}

\begin{document}
\bibliographystyle{unsrt}

\title[Complex magnetic monopoles, geometric phases and quantum evolution  \dots]{Complex magnetic monopoles, geometric phases and quantum evolution in vicinity of diabolic and exceptional points }

%\title[Complex magnetic monopoles and geometric phases  \dots]{Complex magnetic monopoles and geometric phases around diabolic and exceptional points}

\author{Alexander I Nesterov and F. Aceves de la Cruz}

\address{Departamento de F{\'\i}sica, CUCEI, Universidad de Guadalajara, Guadalajara, Jalisco, M\'exico}
\ead{nesterov@cencar.udg.mx, fermincucei@gmail.com}

\begin{abstract}
We consider the geometric phase and quantum tunneling in vicinity of diabolic and exceptional points. We show that the geometric phase associated with the degeneracy points is defined by the flux of complex magnetic monopole. In weak-coupling limit the leading contribution to the real part of geometric phase is given by the flux of the Dirac monopole plus quadrupole term, and the expansion for its imaginary part starts with the dipolelike field. For a two-level system governed by the generic non-Hermitian Hamiltonian, we derive a formula to compute the non-adiabatic complex geometric phase by integral over the complex Bloch sphere. We apply our results to
to study a two-level dissipative system driven by periodic electromagnetic field and show that in the vicinity of the exceptional point the complex geometric phase behaves as step-like function. Studying tunneling process near and at exceptional point, we find two different regimes: coherent and incoherent. The coherent regime is characterized by the Rabi oscillations and one-sheeted hyperbolic monopole emerges in this region of the parameters. In turn with the incoherent regime the two-sheeted hyperbolic monopole is associated. The exceptional point is the critical point of the system where the topological transition occurs and both of the regimes yield the quadratic dependence on time. We show that the dissipation brings into existence of pulses in the complex geometric phase and the pulses are disappeared when dissipation dies out. Such a strong coupling effect of the environment is beyond of the conventional adiabatic treatment of the Berry phase.

%{\em Version: dep-jpa-vf-r.tex, June 22, 2008}

\end{abstract}

\pacs{03.65.Vf, 14.80.Hv, 03.65.-w, 03.67.-a, 11.15.-q}

%\keywords{Berry phase, Dirac monopole, complex geometric phase, quantum tunneling }
%display desired

\maketitle

\section{Introduction}

The geometric phase was first discovered by Berry
\cite{B0} in the context of the adiabatic cyclic evolution of the quantum
system. Later Wilczek and Zee \cite{WZ} generalized the Berry's phase allowing
the transported states to be degenerated, and Aharonov and Anandan
\cite{AA} extended Berry's result to non-cyclic and non-adiabatic variation of
the parameters of the Hamiltonian. However, while notion of the geometric phase for pure states is well defined, the definition of a geometric phase in open quantum systems is still an unsolved problem.

The first important step towards to the consistent description of geometric phase for open systems was done by Garrison and Wright \cite{GW}. They removed the restriction to unitary evolution considering the quantum system when the time evolution in governed by non-Hermitian Hamiltonian. Beginning with the classical works by Weisskopf and Wigner on metastable systems \cite{WW,WW1}, modeling the dissipative quantum systems by effective non-Hermitian Hamiltonian is well known. It was observed that for the system being initially in the metastable state $\psi(0)$, its evolution is described by the effective non-Hermitian Hamiltonian as follows: $\psi(0)\rightarrow \psi(t)= e^{-iH_{ef}t}\psi(0)+$ decay products. Its derivation can be made by separation of the full Hilbert space into the intrinsic discrete part and continua and applying the projection operator technique to eliminate the continuum part
(see e.g. \cite{JMR,RI,RI1}). Since the Garrison and Wright paper has been published, the geometric phase for quantum systems governed by the non-Hermitian Hamiltonian and complex-valued geometric phase effects in dissipative systems were studied by various authors (for discussions and references see, e.g., \cite{GW,BD,B,B1,B2,B3,GXQ,H,H2,HH,KKm,KM,DMT,AMPTV,MS,MH}).

Actually, for correct description of open quantum system a density-matrix approach is required. In particular, when the state of the system is changed negligible during the time scale characterizing the decay of the reservoir correlation function, the Born-Markov approximation is valid, and an open quantum system can be described by the following general master equation written in the Lindblad form ($\hslash =1$) \cite{LG}:
\begin{eqnarray}\label{D1}
\dot\rho= -i[H, \rho] - \frac{1}{2}\sum_{k=1}^{N}\{\Gamma_k^\dagger \Gamma_k
\rho + \rho\Gamma_k^\dagger \Gamma_k -2 \Gamma_k^\dagger \rho\Gamma_k\},
\end{eqnarray}
where $H$ is the Hamiltonian of the uncoupled system and $\Gamma_k$ are
transition operators describing the coupling to the reservoir. The commutator
of the density operator $\rho$ with a Hamiltonian $H$ represents the coherent
part of evolution and the remaining part corresponds to the decoherence process on account of the interaction with an environment. The first analysis of geometric phase based on the master equation has been done by Ellinais {\em et al} \cite{EBD} and Gamliel and Freed \cite{GF}.

To make connection with the non-unitary evolution of quantum system, consider the quantum jump approach to open systems, in which the evolution of system may be separated in a part describing ``no-jump" and spontaneous jumps part as follows. The total evolution time $T$ is divided into a sequence of intervals $\Delta t = T/N$. Then, the jump-free trajectory being governed by the effective non-Hermitian Hamiltonian
\begin{equation}\label{H1}
\widetilde  H = H - \frac{i}{2} \sum_{k=1}^{n} \Gamma_k^\dagger\Gamma_k,
\end{equation}
is interrupted by instantaneous jumps caused by jump operators $\Gamma_k$:
$\rho(t_m) \rightarrow \rho (t_{m+1})=\sum_{k=0}^{n}W_k
\rho(t_m)W_k^\dagger$, where $W_0 = {1\hspace{-.125cm}1}- i\widetilde H \Delta t$, $W_k =\sqrt{\Delta t}\Gamma_k$, for $k=1,2,\dots, n$, and $t_m= m\Delta t$ \cite{CFSV,PK}. Recently, these ideas has been employed to show how the geometric phase can be modified for open systems \cite{CFSV,CFSV1}.

Recent experimental results providing evidence for the `magnetic' monopole in the crystal-momentum space \cite{FNT} and emerging of `fictitious magnetic monopoles' in the anomalous Hall effect of ferromagnetic materials, magnetic superconductors, trapped $\Lambda$-type atoms, anisotropic spin systems, noncommutative quantum mechanics, in ferromagnetic spinor Bose-Einstein condensate, etc. \cite{FNT,ZLS,Br,Hal,FP,SR,MSN,BM}, has been caused a rebirth interest in the Dirac monopole problem.

The `fictitious' monopoles appear in the context of the Berry phase as follows. Assume that for adiabatic driving quantum system the energy levels may cross \cite{B0}. Then, in the commonest case of double degeneracy with two linearly independent eigenvectors, the energy surfaces form the sheets of a double cone, and its apex is called a ``diabolic point'' \cite{BW}. Since for generic Hermitian Hamiltonian the codimension of the diabolic point is three, it can be characterized by three parameters $\mathbf R= (X,Y,Z)$. The eigenstates $|n,\mathbf R \rangle$ give rise to the Berry's connection defined by ${\mathbf A}_n(\mathbf R)= i\langle n,\mathbf R| \nabla_{\mathbf R} |n,\mathbf R \rangle$, and the curvature
$\mathbf B_n = \nabla_{\mathbf R} \times {\mathbf A}_n $ associated
with ${\mathbf A}_n$ is the field strength  of `magnetic' monopole located at
the diabolic point \cite{B0,BD}. The Berry phase $\gamma_n= \oint_{\mathcal C }{\mathbf A}_n \cdot d \mathbf R$ is interpreted as a holonomy associated with the parallel transport along a circuit $\mathcal C$ \cite{SB}.

In opposition to the diabolic point, at which the eigenvalues coincide while the eigenvectors still remain distinct, an exceptional point occurs when the eigenvalues and eigenvectors coalesce. This type of degeneracy is associated with open quantum systems and non-Hermitian physics \cite{CFSV,PK,BHC,B,H1}. The exceptional points have been observed in various physical systems: laser induced ionization of atoms, microwave cavities, ``crystals of light'', in optics of absorptive media, electronic circuits, etc. \cite{LKDP,PBPR,OMAR,DGHH,SHS}

In this paper we consider the geometric phase and tunneling process near and at diabolic and exceptional points. We show that for general non-Hermitian system the geometric phase associated with the degeneracy is described by {\em complex magnetic monopole}. We find that the exceptional point is the bifurcation point of the complex geometric phase in the parameter space and the real part of the latter has a jump discontinuity at the exceptional point. We show that the exceptional point is the critical point of the quantum-mechanical system, where the topological phase transition in the parameter space occurs.

Studying the tunneling process in the vicinity of exceptional point we found two distinct regimes: coherent and incoherent. The coherent tunneling is characterized by the Rabbi oscillations, also known as quantum echoes. We also show that the dissipation brings into existence of pulses in the real part of the geometric phase. Such a strong coupling effect of the environment disappears in the absence of dissipation.

\section{General results on behavior of the eigenvectors at diabolic and
exceptional points}

As known, in parameter space, a set of exceptional points defines a smooth surface of codimension two for symmetric/nonsymmetric complex matrix, codimension one for a real nonsymmetric matrix, and exceptional point's do not exist for real symmetric or Hermitian matrix \cite{Arn}. Let $H(X)$ be a complex $N\times N$ matrix smoothly dependent on $m$ real
parameters $X_i$, $i$ runs from 1 to m.  For $\lambda_k(X)$ being the
eigenvalues of $H(X)$, we denote by $|\psi_k(X)\rangle$ and
$\langle\tilde\psi_k(X)|$ be the corresponding right/left eigenvectors:
\begin{equation}\label{Eq21}
    H|\psi_k\rangle = \lambda_k|\psi_k\rangle, \quad \langle\tilde\psi_k|H=
    \lambda_k\langle\tilde\psi_k|
\end{equation}
Both systems of left and right eigenvectors form a bi-orthogonal basis
\cite{MF}
\begin{eqnarray}
\sum_k\frac{|\psi_k\rangle\langle\tilde\psi_k|}{\langle\tilde\psi_k|\psi_k\rangle}=
1 \quad \langle\tilde\psi_k|\psi_{k}\rangle = 0,\quad k \neq k' \label{Beq}.
\end{eqnarray}
Using the decomposition of unity (\ref{Beq}), we obtain
\begin{eqnarray}\label{Eq22}
|\psi\rangle = \sum_i \alpha_i |\psi_k\rangle, \quad \langle\widetilde\psi|=
\sum_i \beta_i \langle\tilde\psi_i|
\end{eqnarray}
where
\begin{eqnarray}\label{Eq23}
\alpha_i =
\frac{\langle\tilde\psi_i|\psi\rangle}{\langle\tilde\psi_i|\psi_{i}\rangle},
\quad \beta_i =
\frac{\langle\widetilde\psi|\psi_{i}\rangle}{\langle\tilde\psi_i|\psi_{i}\rangle}.
\end{eqnarray}

We assume that exceptional point occurs for some value of parameters $X=X_c$. At the exceptional point the eigenvalues, say $n$ and $n+1$, coalesce: $\lambda_n(X_c)=\lambda_{n+1}(X_c)$, and the corresponding eigenvectors coincide, up to a complex phase, yielding a single eigenvector
$|\psi_{EP}\rangle$. Now, applying (\ref{Beq}) for $k=n$ and $k=n+1$ we find that at the exceptional point the normalization condition is violated,
%\begin{equation}\label{Beq_2}
$\langle\tilde\psi_{EP}|\psi_{EP}\rangle =0.$
%\end{equation}
This leads to the serious consequences for the global behavior of the states on parameter space.

Since at exceptional point both eigenvalues and eigenvectors merge forming a Jordan block, it is convenient to introduce the orthonormal basis of the related invariant 2-dimensional subspace as follows:
\begin{eqnarray}
\langle n|n\rangle =1, \quad \langle n+1|n+1\rangle =1, \quad \langle
n|n+1\rangle
 =0
\end{eqnarray}
Assuming that all other eigenstates are non-degenerate, we find that the set of vectors $\{ |\chi_k \rangle, \langle\tilde\chi_k |, \}$ , where $ |\chi_n
\rangle = |n\rangle,\quad  \langle\tilde\chi_{n+1} |= \langle n+1|$, and
\begin{eqnarray*}
|\chi_k \rangle = \frac{|\psi_k\rangle}{\sqrt{\langle
\tilde\psi_k|\psi_k\rangle}}, \quad \langle\tilde\chi_k | =
\frac{\langle\tilde\psi_k |}{\sqrt{\langle
 \tilde\psi_k|\psi_k\rangle}}, \; {\rm for}\; k\neq n,n+1
\end{eqnarray*}
form the bi-orthonormal basis. Using this basis we expand an arbitrary vector
$\psi$ as
\begin{equation}\label{Exp}
|\psi\rangle =  \sum c_k (X)|\chi_k (X) \rangle
\end{equation}
with the coefficients of expansion defined as $c_k = \langle
\tilde\chi_k|\psi\rangle $.

From the orthogonality condition, one can see that if $|\psi(X)\rangle
\rightarrow |\psi_{EP}\rangle$ while $X \rightarrow X_c$, all coefficients $c_k$ ($k\neq n,n+1$) vanish at exceptional point. Thus in the neighborhood of exceptional point only the terms related to the invariant subspace make substantial contributions and the $N$-dimensional problem becomes effectively two-dimensional \cite{Arn,KMS}. Similar conclusion is valid for the diabolic point, excepting that at the diabolic point the eigenvectors remain distinct.

Now, let $|\psi\rangle$ being the eigenvector of $H$,
\begin{equation}
H|\psi(X)\rangle=\lambda(X) |\psi(X)\rangle,
\end{equation}
belongs two-dimensional invariant subspace defined by the Jordan block. This
implies that in the expansion (\ref{Exp}) for $k\neq n,n+1$ all $c_k=0$, and we
may write
\begin{equation}\label{Exp2}
|\psi\rangle = \alpha |n\rangle + \beta |n+1\rangle
\end{equation}
For $X\neq X_c$, one has two eigenvalues $\lambda_{+}$ and $\lambda_{-}$ and
the corresponding eigenvectors $|\psi_{\pm}\rangle$ are given by
\begin{equation}\label{eqH}
|\psi_{\pm}\rangle = \alpha_{\pm} |n\rangle + \beta_{\pm} |n+1\rangle
\end{equation}

The coefficients $\lambda_{\pm},\alpha_{\pm}$ and $\beta_{\pm}$ are found from
the two-dimensional eigenvalue problem
\begin{equation}\label{eqH1}
\left(
  \begin{array}{cc}
    a_{11} & a_{12} \\
   a_{21} &a_{22} \\
  \end{array}
\right) \left(
  \begin{array}{c}
    \alpha_{\pm} \\
    \beta_{\pm} \\
  \end{array}
\right)
 =\lambda_{\pm}
\left(
  \begin{array}{c}
    \alpha_{\pm} \\
    \beta_{\pm} \\
  \end{array}
\right)
\end{equation}
where the matrix elements are determined as
\begin{eqnarray*}
a_{11} = \langle n|H|n\rangle, \quad &&a_{12} = \langle n|H|n+1\rangle, \\
a_{21} = \langle n+1|H|n\rangle, \quad &&a_{22} = \langle n+1|H|n+1\rangle
\end{eqnarray*}
It is convenient to introduce the following notations:
\begin{eqnarray}
\lambda_0= \frac{a_{11}+a_{22}}{2},\quad  X= \frac{a_{12}+a_{21}}{2}, \\
Y= \frac{a_{21}-a_{12}}{2i}, \quad  Z= \frac{a_{11}-a_{22}}{2},
\end{eqnarray}
Then Eq.(\ref{eqH1}) reads
\begin{equation}\label{eqH2}
\left(
  \begin{array}{cc}
    \lambda_0 + Z & X-iY \\
   X+iY & \lambda_0 - Z\\
  \end{array}
\right) \left(
  \begin{array}{c}
    \alpha_{\pm} \\
    \beta_{\pm} \\
  \end{array}
\right)
 =\lambda_{\pm}
\left(
  \begin{array}{c}
    \alpha_{\pm} \\
    \beta_{\pm} \\
  \end{array}
\right)
\end{equation}
Solving the characteristic equation for (\ref{eqH2}), we obtain
\begin{equation}\label{L}
\lambda_{\pm}= \lambda_{0} \pm \sqrt{X^2 + Y^2 + Z^2}
\end{equation}
The eigenvalues coalescence at the point $R=0$, that yields the diabolic point if
$X=Y=Z=0$, and the exceptional point otherwise. The detailed study of two-dimensional problem
will be presented in the following sections.

\section {Degeneracy, geometric phases and complex `magnetic'
 monopoles}

Following \cite{GW}, let us consider the time dependent Schr\"odinger equation and its adjoint equation:
\begin{eqnarray}\label{S1}
i\frac{\partial }{\partial t}|\Psi(t)\rangle = H(X(t))|\Psi(t)\rangle, \\
-i\frac{\partial }{\partial t}\langle\widetilde\Psi(t)| =
\langle\widetilde\Psi(t)|H(X(t))\label{S2},
\end{eqnarray}
where $H$ is the non-Hermitian Hamiltonian.

Let $\langle\tilde\psi_n(X)|$ and $|\psi_n(X)\rangle$ be left (right)
eigenstates corresponding to the eigenvalue $E_n$, then in adiabatic
approximation the complex geometric phase is given by \cite{GW,B1,B2}
\begin{eqnarray}\label{Eq10}
\fl \gamma_n = i\oint_C
\frac{\langle\tilde\psi_n(X)|d\psi_n(X)\rangle}{\langle\tilde\psi_n(X)|\psi_n(X)\rangle}
\end{eqnarray}
generalizing Berry's result to the dissipative case. Further we assume that the instantaneous eigenvectors form the bi-orthonormal basis, $\langle\tilde\psi_m|\psi_n\rangle =\delta_{mn}$. This can alter the geometric phase (\ref{Eq10}) up to the topological contribution $\pi n$, $n\in \mathbb Z$ \cite{MKS,GRS}.

For a non-Hermitian Hamiltonian, validity of the adiabatic approximation is defined by the following condition:
\begin{equation}\label{Eq24}
    \sum_{m\neq n}\bigg|\frac{\langle\tilde\psi_m|\partial H/\partial
    t|\psi_n(X)\rangle}{(E_m - E_n)^2}\bigg|\ll 1
\end{equation}
This restriction is violated nearby the degeneracies related to any of  diabolic point or exceptional point, where the eigenvalues coalesce.

Since the adiabatic approach cannot be applied in the neighborhood of
degeneracy, we will consider non-adiabatic generalization of Berry's phase introduced by Aharonov and Anandan \cite{AA} and extended by Garrison and Wright to the non-Hermitian systems as follows \cite{GW}. Let an adjoint pair $\{|\Psi(t)\rangle, \langle\widetilde\Psi(t)|\}$ being a solution of Eqs. (\ref{S1}), (\ref{S2}) satisfies the following condition
\begin{eqnarray}\label{Eq25}
&|\Psi(T)\rangle = \exp(i\varphi)|\Psi(0)\rangle, \\
&\langle\widetilde\Psi(T)|= \exp(-i\varphi)\langle\widetilde\Psi(0)|,
\end{eqnarray}
where $\varphi$ is complex, and $\{|\chi(t)\rangle, \langle\tilde\chi(t)|\}$ be a modified adjoint pair such that
\begin{eqnarray}\label{Eq26}
&|\chi(t)\rangle = \exp(if(t))|\Psi(t)\rangle, \\
&\langle\tilde\chi(t)|= \exp(-if(t))\langle\widetilde\Psi(t)|,
\end{eqnarray}
where $f(t)$ is any function satisfying $f(t+T) - f(t)= \varphi(t)$. The total
phase $\varphi$ calculated for the time interval $(0,T)$ may be written as
$\varphi = \gamma + \delta$, where the  ``dynamical phase'' is given by
\begin{equation}\label{Eq27h}
 \delta = -\int_0^T \langle\tilde\chi(t)|H|\chi(t)\rangle dt.
\end{equation}
and for the geometric phase $\gamma$ one has.
\begin{equation}\label{Eq27}
 \gamma = i\int_0^T \langle\tilde\chi(t)|\frac{\partial}{\partial
 t}\chi(t)\rangle dt
\end{equation}
This yields the connection one-form and the curvature two-form as follows
\cite{AS}:
\begin{equation}\label{Eq11}
A= i\langle\tilde\chi|d\chi\rangle, \quad F=dA.
\end{equation}
Note that real part of  geometric phase (\ref{Eq27}) besides of the usual Berry phase, contains the contribution of environment. Its imaginary part changes the amplitude of the density matrix and implies mixture of the initially pure states.

Geometric phase $\gamma$ for an arbitrary quantum evolution can be obtained, also, from the total phase $\gamma_t$ by subtracting the dynamical phase $\gamma_d$ \cite{MS}:
\begin{equation}\label{GP3}
\gamma= \gamma_t -\gamma_d,
\end{equation}
where
\begin{eqnarray}
\label{GP4}
  \gamma_t = \arg\langle \Psi(0)|\Psi(t) \rangle, \quad {\rm and} \quad
  \gamma_d = -i \int_0^t \langle \Psi(t)|\frac{d}{dt}|\Psi(t) \rangle dt .
\end{eqnarray}

We adopt and generalize this definition of the geometric phase  for non-Hermitian quantum evolution as follows (see also \cite{MKS}):
\begin{eqnarray}
\label{GP5}
\gamma = \frac{i}{2} \ln\frac{\langle\widetilde \Psi(t)|\Psi(0) \rangle}{\langle\widetilde \Psi(0)|\Psi(t) \rangle} + i \int_0^t \langle \widetilde\Psi(t)|\frac{d}{dt}|\Psi(t) \rangle dt
\end{eqnarray}
As can be observed, (\ref{GP5}) yields gauge invariant definition of the geometric phase with respect of gauge transformations:
\begin{equation}\label{GT}
|\Psi \rangle \rightarrow e^{i\alpha} |\Psi \rangle , \; \langle\widetilde \Psi|  \rightarrow e^{-i\alpha} \langle\widetilde \Psi|, \; \alpha \in \mathbb C.
\end{equation}

\subsection{Two-level system and `magnetic' monopoles}

As has been mentioned before, in vicinity of degeneracy point behavior of $N$-dimensional system can be described by the effective two-dimensional quantum system. In what follows, we consider in detail the complex geometric phase associated with the generic non-Hermitian Hamiltonian:
\begin{equation}\label{eqH2a}
H=\left(
  \begin{array}{cc}
    \lambda_0 + Z & X-iY \\
   X+iY & \lambda_0 - Z\\
  \end{array}
\right), \quad X,Y,Z \in \mathbb C
 \end{equation}
$X,Y,Z \in \mathbb C$ being complex parameters.

For the Hamiltonian (\ref{eqH2a}) the exceptional point is determined by equation
\begin{equation}\label{Eq9}
X^2 +Y^2 + Z^2=0,
\end{equation}
and defines a hypersurface of complex codimension 1 in $\mathbb C^3$, which also can considered as a smooth surface of codimension 2 in 6-dimensional real space. Note, that the diabolic point being just a point in 3-dimensional complex space $\mathbb{C}^3$, is located at the origin of coordinates.

The solution of the eigenvalue problem
\begin{equation}\label{exceptional point1}
H|u\rangle =\lambda|u\rangle, \quad \langle \tilde u| H =\lambda \langle \tilde
u|,
\end{equation}
where $|u\rangle$ and $\langle \tilde u|$ are the right and left eigenvectors,
respectively, is given by
\begin{equation}
\lambda_{\pm} = \lambda_0 \pm R, \label{EigL}
\end{equation}
where $R = {(X^2 +Y^2 + Z^2)}^{1/2}$. The right and left eigenvectors are found to be
\begin{eqnarray}
&|u_{+}\rangle = \left(\begin{array}{c}
                  \cos\frac{\theta}{2}\\
                  e^{i\varphi}\sin\frac{\theta}{2}
                  \end{array}\right),
\langle \tilde u_{+}| = \bigg(\cos\frac{\theta}{2},
e^{-i\varphi}\sin\frac{\theta}{2}\bigg)  \label{r}\\
&|u_{-}\rangle = \left(\begin{array}{c}
-e^{-i\varphi}\sin\frac{\theta}{2}\\
\cos \frac{\theta}{2} \end{array}\right), \; \langle \tilde u_{-}|
=\bigg(-e^{i\varphi}\sin\frac{\theta}{2}, \cos\frac{\theta}{2}\bigg) \label{l}
\end{eqnarray}
where
\begin{eqnarray}\label{Eq15}
\cos\frac{\theta}{2}= \sqrt{\frac{R+Z}{2R}}, & \quad
\sin\frac{\theta}{2}=\sqrt{\frac{R-Z}{2R}}, \\
e^{i\varphi} = \frac{X + iY}{\sqrt{R^2 -Z^2}}&,\quad e^{-i\varphi} = \frac{X - iY}{\sqrt{R^2 -Z^2}},
\label{Eq15a}
\end{eqnarray}
and $\theta, \varphi$ are the complex angles of the complex spherical coordinates:
\begin{eqnarray}\label{Eq10a}
X=R \sin\theta \cos\varphi, \;
Y=R \sin\theta \sin\varphi, \;
Z= R \cos\theta
\end{eqnarray}
Finally, for $R\neq 0$ the following relationships hold
\begin{equation}
\langle \tilde u_{\pm}|u_{\mp}\rangle = 0, \quad \langle \tilde
u_{\pm}|u_{\pm}\rangle = 1 \label{orthogen}
\end{equation}

As seen from Eq.(\ref{EigL}), the coupling of eigenvalues $\lambda_{+}$ and
$\lambda_{-}$ occurs when $X^2 +Y^2 + Z^2 =0$. This yields two cases.
The first one, defined by $\theta=0, \; \varphi =0$, yields two linearly independent eigenvectors. The point of coupling is known as the diabolic point, and we obtain
\begin{eqnarray}
|u_{+}\rangle = \left(\begin{array}{r}
                  1\\
                  0
                  \end{array}\right),\;
\langle \tilde u_{+}| = (1, 0), \;\;
|u_{-}\rangle = \left(\begin{array}{c}
0\\
1 \end{array}\right), \;
\langle \tilde u_{-}| =(0, 1) \label{left},
\end{eqnarray}
The second case is characterized by coupling of eigenvalues and merging of eigenvectors, as well. The degeneracy point is known as the exceptional point, and we have: $|u_{+}\rangle= e^{i\kappa}|u_{-}\rangle$ and $\langle \tilde u_{+}|=e^{-i\kappa}\langle \tilde u_{-}|$, where $\kappa \in \mathbb C$ is a complex phase. Hence, the violation of the normalization condition (\ref{orthogen}) is occurred at the exceptional point, and we have $\langle \tilde u_{\pm}|u_{\pm}\rangle = 0$.

Let us assume that the exceptional point is given by  $\mathbf R_0 = (X_0,Y_0,Z_0)$. Then, if $Z_0 \neq 0$, using Eqs. (\ref{Eq15}) -- (\ref{Eq10a}) we obtain
\begin{eqnarray}
\tan\frac{\theta_0}{2} =\pm i, \quad
e^{2i\varphi_0} = \frac{X_0 + iY_0}{X_0 - iY_0},
\label{Eq15b}
\end{eqnarray}
and, thus, at the exceptional point $\Im\theta \rightarrow \pm \infty$. In turns, if $Z_0=0$, we obtain $X_0 = \pm iY_0$. This implies $\theta_0= \pi/2$, and $\Im \varphi \rightarrow \pm \infty$ at the exceptional point.

Inserting formulae (\ref{r}) - (\ref{l}) for $|u_{\pm}\rangle$ and  $\langle
\tilde u_{\pm}|$ into (\ref{Eq11}), we obtain the connection one-form as
\begin{eqnarray}\label{Eq12}
    A=  q(1 -\cos\theta)d\varphi
\end{eqnarray}
where $q= \mp 1/2$ and  upper/lower sign corresponds to $|u_{\pm}\rangle$, respectively. The related curvature two-form reads
\begin{eqnarray}\label{Eq14}
 F=dA=q\sin\theta\; d\theta \wedge d\varphi, \; \theta, \varphi \in \mathbb C,
\end{eqnarray}
and in the complex Cartesian coordinates the  connection one-form and the curvature two-form can be written as
\begin{eqnarray}\label{Eq16}
A = \frac{ q(XdY-YdX)}{R(R + Z)} ,\\
F = \frac{q}{R^3}\varepsilon_{ijk} X^k dX^i\wedge dX^j.
\label{Eq16a}
\end{eqnarray}
The obtained formulae describe complex ``magnetic monopole" with a charge $q$ and the field $B= \ast F$ given by
\begin{equation}\label{Eq4a}
\mathbf B = q\frac{\mathbf R}{R^3},
\end{equation}
where $\mathbf R = (X,Y,Z)$, and $X,Y,Z \in \mathbb C$. The field of the monopole can be written as $B^i = -\partial \Phi/\partial X^i$, where the potential $\Phi = q/R$.

Computation of geometric phase yields
\begin{equation}\label{GP1}
\gamma = \oint_{\mathcal C} A,
\end{equation}
where integration is performed over the contour $\mathcal C$ on the complex
sphere $S^2_c$. Applying the Stokes theorem we obtain
\begin{equation}\label{GP2}
\gamma = \int_{\Sigma} F = q\Omega(\mathcal C),
\end{equation}
where $\Sigma$ is a closed surface with the boundary $\mathcal C= \partial \Sigma $,
and $\Omega(\mathcal C)$ is the complex solid angle subtended
by the contour $\mathcal C$.

Generally, the complex magnetic monopole is emerged in quantum mechanical systems admitting $SL(2,C)$ symmetry. In particular case of $SO(3)$ symmetry the related degeneracy is referred as the diabolic point, and the formula (\ref{Eq4a}) reproduces the classical Berry result on two-fold degeneracy in parameter space \cite{B0}. For the exceptional point the field of the corresponding ``monopole" represents a complicated topological charge rather than a point-like magnetic charge. In what follows we discuss two particular applications: {\em Hyperbolic monopole} and {\em Complex Dirac monopole}.

\subsubsection*{Hyperbolic monopole.}

Let us consider the following non-Hermitian Hamiltonian:
\begin{eqnarray}\label{Eq17}
H=\left(
  \begin{array}{cc}
    \lambda_0+ iz & x-iy \\
    x+iy & \lambda_0- iz \\
  \end{array}
\right), \quad x,y,z \in \mathbb R
\end{eqnarray}
The eigenvalues of the Hamiltonian given by $
\lambda_{\pm}= \lambda_{0} \pm R$, where $R={( x^2+ y^2 -z^2)}^{1/2}$, coalesce at the point $R=0$. In addition, the exceptional point is represented by the double cone with the apex at the origin of coordinates, and the diabolic point is just located at the origin of coordinates.

Applying (\ref{Eq16}) - (\ref{Eq16a}), we obtain
\begin{eqnarray}\label{Eq16b}
A = \frac{ q(xdy-ydx)}{R(R + iz)}, \; {\rm and} \;
F = \frac{iq}{R^3}\varepsilon_{ijk} x^k dx^i\wedge dx^j,
\end{eqnarray}
 This yields
\begin{equation}\label{Eq4h}
\mathbf B = iq\frac{\mathbf R}{R^3}, \quad \mathbf R = (x,y,z)
\end{equation}
Hence, the field $\mathbf B$ can be written as $\mathbf = - \nabla \Phi$, where $\Phi=- iq/R$.

For $R^2> 0$,  the surface defined by $R= \rm const$ is the one-sheeted hyperboloid, which can be considered also as the coset class $SU(1,1)/{\mathbb R} \sim SO(2,1)/SO(1,1)$ \cite{LA, Jack}. Introducing the inner coordinates $(\theta, \varphi)$ as
\begin{eqnarray}
\label{HM}
x = R\cosh \theta\cos\varphi, \;\;  y = R\cosh \theta\sin\varphi, \;\;
z = R\sinh \theta,
\end{eqnarray}
where $-\infty \leq \theta \leq \infty$, $0\leq \varphi < 2\pi$, we obtain
\begin{eqnarray}\label{Eq19}
A=  q(1 -i \sinh\theta)d\varphi\; \; F = -iq\cosh\theta\; d\theta \wedge d\varphi \label{Eq20}
\end{eqnarray}
The obtained one-sheeted hyperbolic monopole carries the imaginary total charge $iq$ and, in contrast to the point-like Dirac monopole, has the singularity on the surface of the double cone identified with the exceptional point.

For  $R^2< 0$, the surface characterized by $z^2 - x^2 -y^2= \rm const$ is the two-sheeted hyperboloid. A convenient parametrization is given by
\begin{eqnarray}
\label{HM1}
x =\tilde R\sinh \theta\cos\varphi, \;\;  y =\tilde R\sinh \theta\sin\varphi, \;\; z = \tilde R\cosh \theta \; \; (z> 0) \\
x =\tilde R\sinh \theta\cos\varphi, \;\;  y =\tilde R\sinh \theta\sin\varphi, \;\; z = - \tilde R\cosh \theta \; \; (z < 0)
\end{eqnarray}
where $0 \leq \theta \leq \infty$, $0\leq \varphi < 2\pi$, $\tilde R= (z^2 - x^2 -y^2)^{1/2}$, and we preserve the same notations for the angular coordinates, as above. The relevant coset class is $SU(1,1)/U(1) \sim SO(2,1)/SO(2)$, and once return to (\ref{Eq16b}) and (\ref{Eq4h}), we obtain
\begin{eqnarray}\label{Eq19a}
A^{+}=  q(1-  \cosh\theta)d\varphi,\; \; F^{+} = - q\sinh\theta\; d\theta \wedge d\varphi \; \; (z>0), \\
A^{-}=  q(1+  \cosh\theta)d\varphi,\; \; F^{-} =  q\sinh\theta\; d\theta \wedge d\varphi \; \;(z < 0).
\end{eqnarray}
Hence, the obtained monopole carries a real total charge given by $-q$. Note, that the two-sheeted hyperbolic monopole has been already appeared in the literature in connection with the geometric phase (see, e.g.\cite{LA, Jack,VL}). Moreover, as has been pointed out by Jackiw \cite{Jack}, this is a topologically trivial case, and the curvature may be removed by a globally well-defined canonical transformation.

\begin{figure}[tbh]
%\begin{minipage}[]{8.5cm}
\scalebox{0.325}{\includegraphics{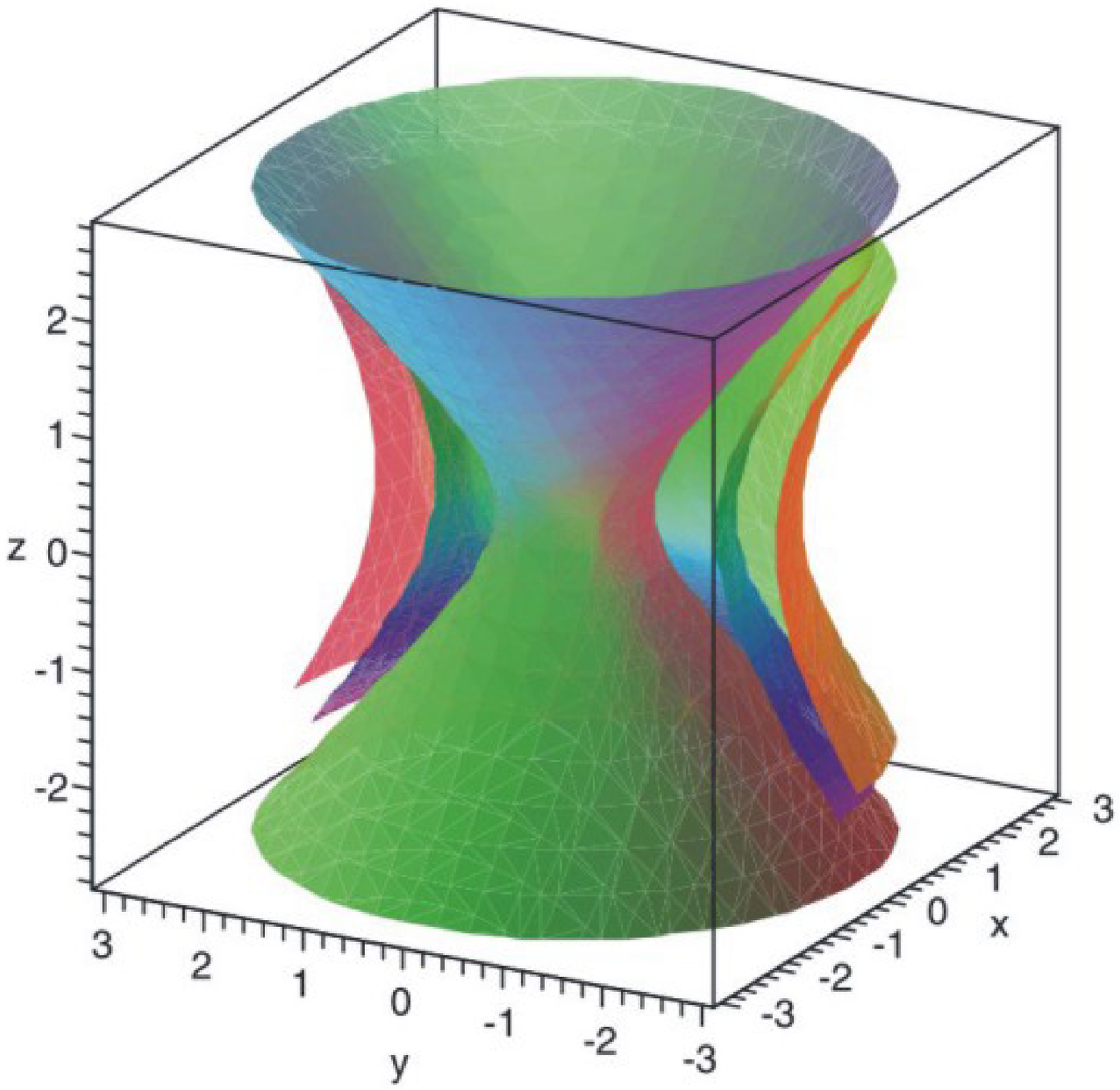}} \scalebox{0.325}{\includegraphics{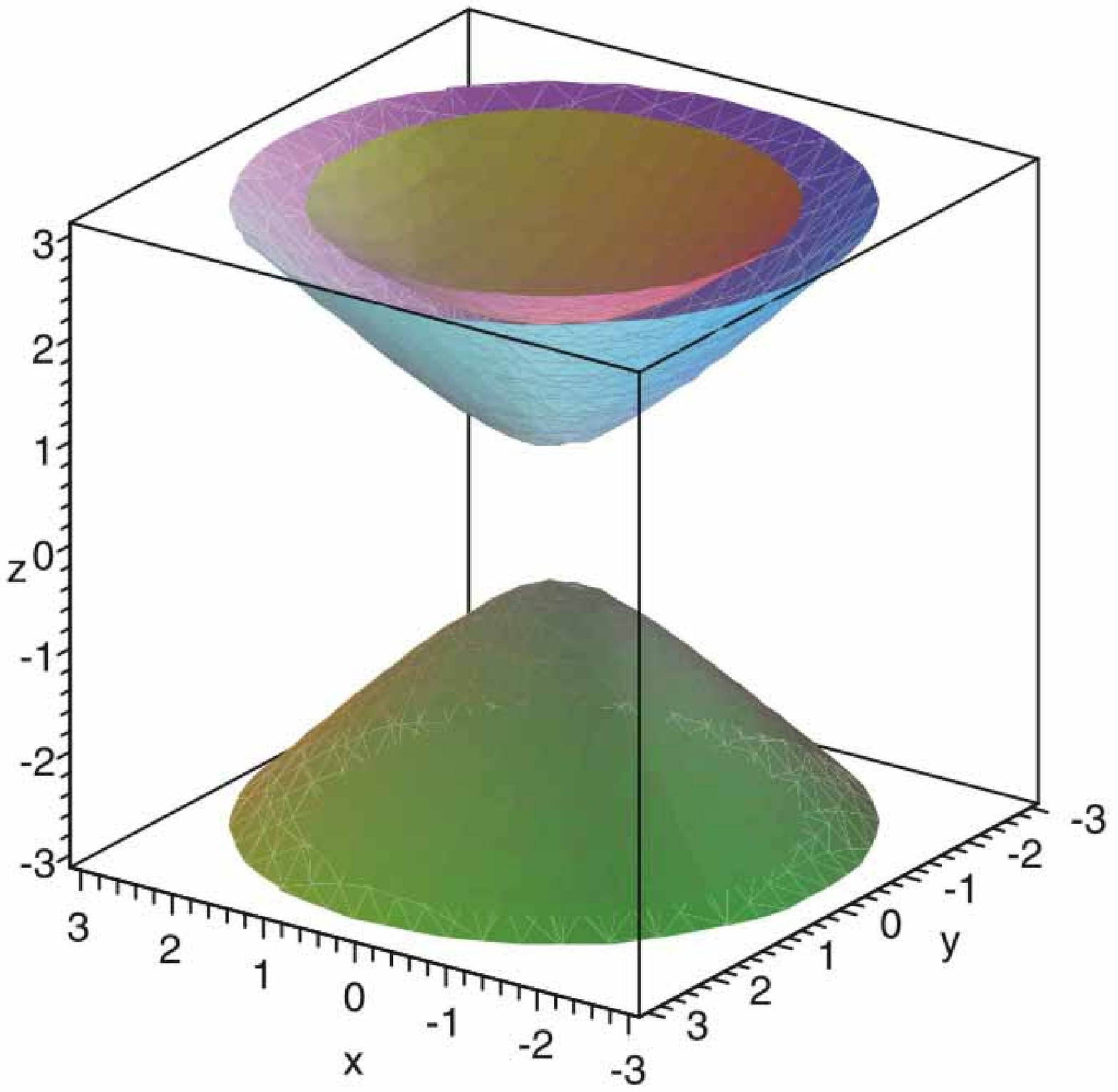}}
\caption{Hyperbolic monopole. Left panel: $R^2> 0$, one-sheeted hyperbolic monopole. The surfaces $\Im \Phi = \rm const$ are depicted. Right panel: $R^2 < 0$, two-sheeted hyperbolic monopole. The surfaces $\Re \Phi = \rm const$ are presented. The exceptional point is realized as the double-cone (not plotted).
\label{HMC}}
%\end{minipage}
\end{figure}

As can be easily shown, in the case of the one-sheeted monopole the corresponding potential $\Phi$ is imaginary, and for the two-sheeted hyperbolic monopole $\Phi$ is a real function. In Fig. \ref{HMC} the surfaces of $\Im \Phi = \rm const$ and $\Re \Phi = \rm const$ related to one-sheeted and two-sheeted hyperbolic monopole, respectively, are depicted.

The hyperbolic monopoles are appeared in a wide class of physical systems admitting $SO(2,1)$ invariance (for discussion see, e.g. \cite{LA, Jack,VL} and references therein). For instance,  the hyperbolic monopole is emerged  in a two-level atom interacting with an electromagnetic field  (see Section 4.1.3)

\subsubsection*{Complex Dirac monopole.}

Let us consider the non-Hermitian Hamiltonian written as
\begin{eqnarray}\label{Eq17a}
H=\left(
  \begin{array}{ll}
    \lambda_0+ z -i\varepsilon & x-iy \\
    x+iy & \lambda_0- z +i\varepsilon \\
  \end{array}
\right), \quad x,y,z \in \mathbb R .
\end{eqnarray}
The computation of the `magnetic' field  $\mathbf B$ yields
\begin{equation}\label{Eq4}
\mathbf B =\frac{q\mathbf R}{R^3}, \quad \mathbf R = (x,y,z-i\varepsilon),
\end{equation}
where $R=({ x^2+ y^2 +z^2 -\varepsilon^2 - 2i\varepsilon z})^{1/2}$. The exceptional point obtained as the solution of equation $$ x^2+ y^2 +z^2 -\varepsilon^2 + 2i\varepsilon z=0,$$ is the circle of the radius $\varepsilon$ a the plane $z=0$.

\begin{figure}[tbh]
\scalebox{0.325}{\includegraphics{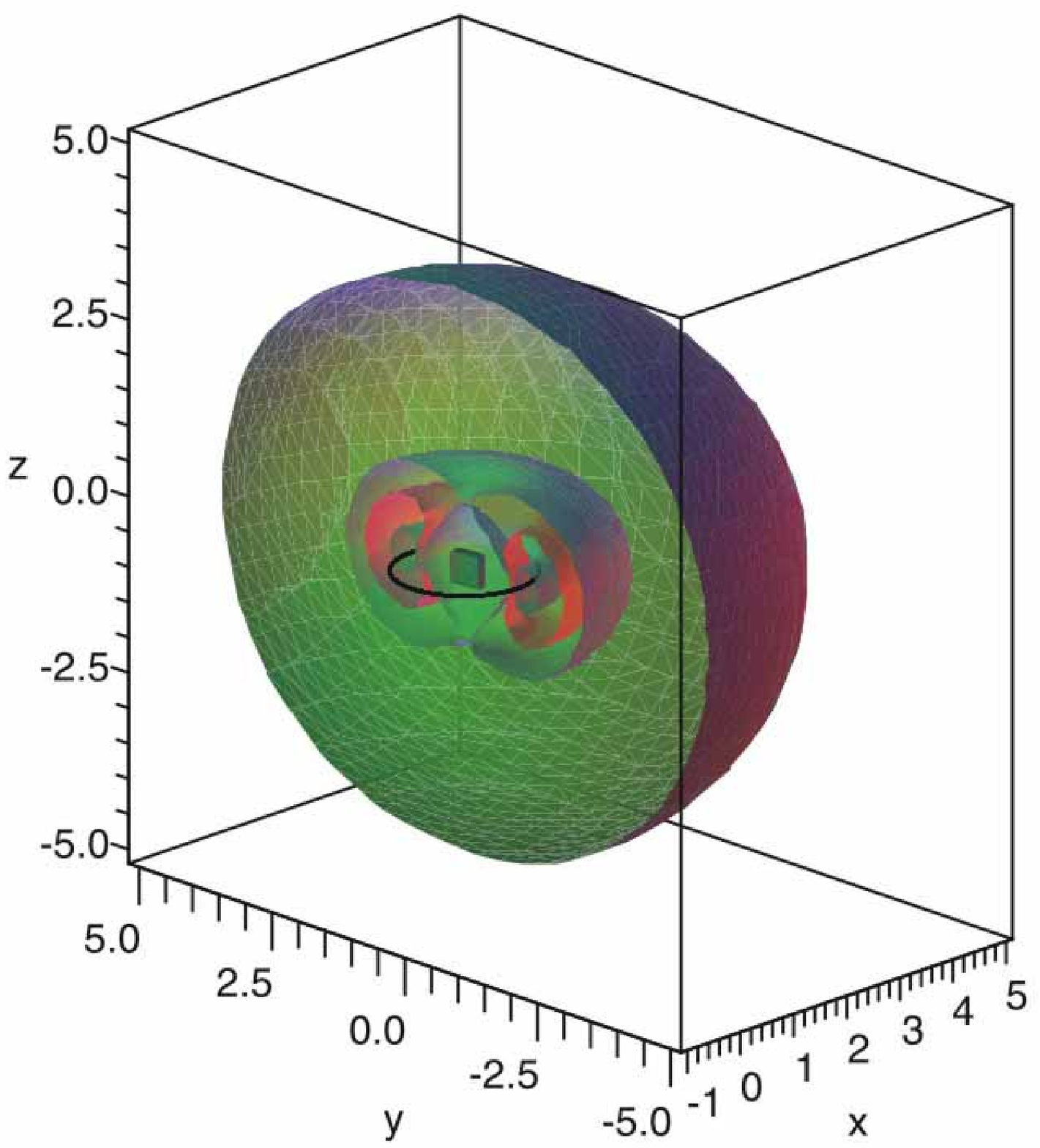}}
\scalebox{0.325}{\includegraphics{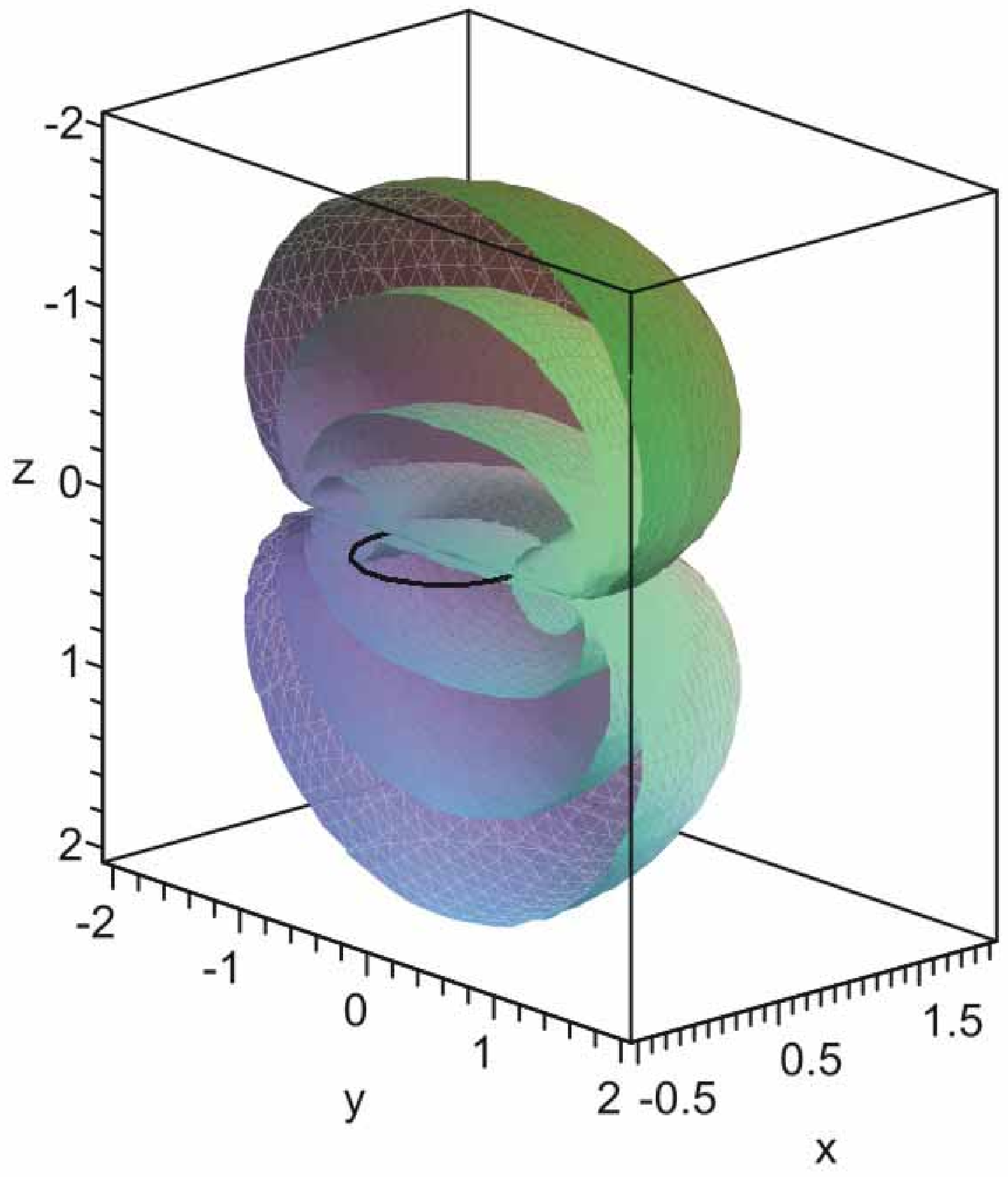}}
\caption{Complex Dirac monopole. The
surfaces $\Re \Phi = \rm const$ (left) and $\Im \Phi = \rm const$ (right) are plotted. The exceptional point is depicted by the circle of the radius
$r=1 \;(\varepsilon = 1)$.  \label{HME}}
\end{figure}

The field of the monopole, $\mathbf B =-\nabla \Phi$, is defined by the complex potential $\Phi =q/R$, depicted in Fig. \ref{HME}.
Setting $r= ({ x^2+ y^2 +z^2})^{1/2}$ and using the real spherical coordinates $(r,\alpha,\beta)$, we have $R= ({r^2 - 2i\varepsilon r \cos\alpha-\varepsilon^2 })^{1/2}$. Then, we may expand the potential $\Phi$ as follows:
\begin{eqnarray}\label{Eq28}
\Phi=\frac{q}{\sqrt{r^2 - 2i\varepsilon r\cos\alpha-\varepsilon^2 }} =
q\sum_{l=0}^{\infty}\frac{(i\varepsilon)^l}{r^{l+1}}P_l(\cos\alpha)
\end{eqnarray}
For $r \gg \varepsilon$, this yields
\begin{eqnarray}\label{Eq29b}
\Phi = \frac{q}{r} + i\frac{p \cos\alpha}{r^2}- \frac{Q(3\cos^2\alpha -1)}{2r^3}
+ \dots ,
\end{eqnarray}
where $q$ is the monopole charge, $p=q\varepsilon$ is the
dipole moment, and $Q=q\varepsilon^2$ is the quadrupole moment.

The geometric phase of the ground state is found to be
\begin{eqnarray}\label{Eq29a}
\gamma = q\oint_{\mathcal C}\bigg(1- \frac{r\cos\alpha-i\varepsilon}{\sqrt{r^2-
2i\varepsilon r \cos\alpha-\varepsilon^2}}\bigg)d\beta
\end{eqnarray}
Using multipole expansion (\ref{Eq29b}), we obtain for $\gamma$ the following
expression:
\begin{eqnarray}\label{Eq29}
\gamma = \gamma_M +i \oint_{\mathcal C} \frac{p \sin^2\alpha}{r} d\beta -
\oint_{\mathcal C} \frac{3Q \sin^2\alpha \cos\alpha}{2r^2} d\beta+\dots ,
\end{eqnarray}
where $\gamma_M= q\oint_{\mathcal C}(1-\cos\alpha)d\beta$ is the contribution
of the Dirac monopole at the origin, the second term describes the dipole
contribution to the imaginary part of the geometric phase and the third term
the quadrupole contribution to its real part.

Let us consider the closed curve
$\mathcal C$ parameterized by $\beta$ with the complex angle
$\theta=\rm const$. Then for q=1/2 the geometric phase (\ref{Eq29a}) becomes
\begin{equation}\label{Eq30}
\gamma= \pi\Bigg( 1- \frac{z - i\varepsilon}{\sqrt{\rho^2+ (z -
i\varepsilon)^2}}\Bigg)
\end{equation}
where $\rho =(x^2 +y^2)^{1/2}$.

A complex Dirac monopole and related complex Berry's phase appear in wide class of open systems, where the Hamiltonian
$$\tilde H= \mathbf B(t)\cdot \boldsymbol \sigma-
\frac{i}{2}\Gamma^\dagger \Gamma
$$
includes spontaneous decay $\Gamma = \sqrt{\varepsilon}\sigma_{-}$ as a source
of decoherence (see, e.g., \cite{MH} and references therein). For instance, it emerges in a two-level atom driven by periodic electromagnetic field $\mathbf E(t)= \Re (\boldsymbol{\mathcal  E}(t)\exp(i\nu t))$, with $\boldsymbol {\mathcal E}(t)$ being slowly varied, as follows. In the rotating wave approximation, after removing the explicit time dependence of the Hamiltonian with a suitable nonunitary transformation, the Schr\"odinger equation reads \cite{GW}
\begin{eqnarray}\label{Sch1}
i\frac{\partial }{\partial t}\left(
  \begin{array}{c}
    u_1 \\
   u_2 \\
  \end{array}
\right) =
\frac{1}{2}\left(
  \begin{array}{cc}
    \Delta -i\delta & 2 V^\ast\\
    2V & -  \Delta +i\delta \\
  \end{array}
  \right)
\left(
  \begin{array}{c}
    u_1 \\
    u_2 \\
  \end{array}
\right)
\end{eqnarray}
where $\delta =(\gamma_a -\gamma_b)/2$, $\gamma_a, \;\gamma_b$ being decay rates for upper and lower levels
respectively; $\Delta= \omega_0 - \nu$, $\omega_0 = (E_a-E_b)$,
$V=\boldsymbol\mu \cdot \boldsymbol{\mathcal E}$, and $\boldsymbol \mu$ is the electric dipole moment. To compare the geometric phase (\ref{Eq30}) with that found in \cite{GW} we set $x= \Re V(t), \; y= \Im V(t), \; z= \Delta/2$ and
$\varepsilon= \delta/2$. Then the geometric phase (\ref{Eq30}) being written as
\begin{equation}\label{Eq30a}
\gamma= \pi\bigg( 1- \frac{\Delta - i\delta}{\sqrt{|2V_0|^2+ (\Delta -
i\delta)^2}}\bigg).
\end{equation}
 coincides with the result obtained by Garrison and Wright \cite{GW} .

\subsubsection*{ Remark.}
Comparing (\ref{HR}) with Eqs. (\ref{Eq17}) and (\ref{Eq17a}), we conclude that both the Dirac complex monopole and hyperbolic monopole can be realized in the four-dimensional parameter space $x,y,z,\varepsilon \in \mathbb R^4 $. The brief classification of the monopole structure is given in the Table 1.
\begin{table}[htb]
\label{Tb}
\begin{center}
\begin{tabular}{|l|l|}  \hline \hline
 \multicolumn{2}{|c|} {\bfseries Monopole classification} \\ \hline
   Dirac monopole &  $\varepsilon  = 0$ \\
\hline
  Complex monopole &  $\varepsilon  \neq 0$  \\
 \hline
  Complex Dirac monopole &   $\varepsilon \neq 0$, $\varepsilon =\rm const$ \\
 \hline
  One-sheeted hyperbolic monopole &  $\varepsilon  \neq 0$, $z = 0$,  $x^2 + y^2 - \varepsilon^2 >0$ \\
 \hline
 Two-sheeted hyperbolic monopole &  $\varepsilon  \neq 0$, $z = 0$,   $x^2 + y^2 - \varepsilon^2 < 0$\\
  \hline \hline
   \end{tabular}
     \caption{Monopole structure of the two-level dissipative system. }
       \end{center}
 \end{table}

\section{Geometric phase and quantum evolution in vicinity of diabolic and exceptional points}

Since the adiabatic approach cannot be applied in the neighborhood of degeneracy, here we consider non-adiabatic generalization of the complex Berry phase. Let $|u(t)\rangle$ and $\langle\tilde u(t)|$ be solutions of the Schr\"odinger equation and its adjoint equation:
\begin{eqnarray}
i\frac{\partial }{\partial t}| u(t)\rangle = H| u(t)\rangle, \\
-i\frac{\partial }{\partial t}\langle\tilde u(t)| =
\langle\widetilde u(t)|H
\end{eqnarray}
where we assume as usual the normalization condition $\langle\tilde u(t)|u(t)\rangle =1$. For an arbitrary evolution of the non-hermitian quantum system the complex geometric phase $\gamma = \gamma_t - \gamma_d$ is given by Eq. (\ref{GP5}), and we have
\begin{eqnarray}
\label{GP5a}
\gamma = \frac{i}{2} \ln\frac{\langle\tilde  u(t)| u(0) \rangle}{\langle\tilde  u(0)| u(t) \rangle} + i \int_0^t \langle \tilde u(t)|\dot u(t) \rangle dt.
\end{eqnarray}

This result can be adopted  to calculate the geometric phase over the complex Bloch sphere as follows. Let $\mathbf n (t) =(\sin \alpha\cos\beta,\sin \alpha\sin\beta, \cos\alpha)$ be a unit complex Bloch vector defined as $\mathbf n(t)= \langle\tilde u(t)|\boldsymbol\sigma|u(t)\rangle$.
One can observe that the Bloch vector satisfies the following equation:
\begin{eqnarray}\label{B1}
    d \mathbf n/dt= \mathbf\Omega \times \mathbf n, \quad\boldsymbol\Omega(t) = {\rm Tr}(H(t)\boldsymbol\sigma),
\end{eqnarray}
and, as shown in the Appendix A, the complex geometric phase can be written as
\begin{eqnarray}\label{Eq28b}
\fl \gamma = -\frac{1}{2}\int_0^\tau (1- \cos\alpha)\dot\beta\, dt + \arctan\frac{\sin(\beta_f - \beta_i)}{\cot(\alpha_f/2)\cot(\alpha_i/2) + \cos(\beta_f - \beta_i)}.
\end{eqnarray}
The integration is performed along the unique curve $\mathbf n(t)$ on the unit sphere $S^2_c$, joining the initial point $\mathbf n(0) =\mathbf n_i = (\sin \alpha_i\cos\beta_i,\sin \alpha_i\sin\beta_i, \cos\alpha_i)$ and the final point $\mathbf n(\tau) = \mathbf n_f = (\sin \alpha_f\cos\beta_f,\sin \alpha_f\sin\beta_f, \cos\alpha_f)$.

Under a cyclic quantum evolution with the period $T$ the Bloch vector describes a closed curve $\mathcal C$ on the complex two-dimensional sphere $S^2_c$, and we have $\mathbf n(t+T)= \mathbf n(t)$. The associated complex geometric phase, being half of the complex solid angle enclosed by $\mathcal C$,
\begin{equation}\label{Eq27g}
 \gamma = -\frac{1}{2}\oint (1 -\cos\alpha) d\beta  = -\frac{1}{2} \Omega(\mathcal C),
\end{equation}
is known as the complex Aharonov-Anandan phase \cite{GW,CWZL}.

Consider a generic non-Hermitian Hamiltonian
\begin{equation}\label{eqH3}
H= \frac{\lambda_0}{2} {1\hspace{-.17cm}1}+ \frac{1}{2}{\boldsymbol { \Omega}}\cdot \boldsymbol \sigma,
\end{equation}
where  ${1\hspace{-.15cm}1}$ denotes the identity operator. Let $|u_i\rangle$ be a given initial state, then the solution of the Schr\"odinger equation (\ref{Eq28b}) can be written as $|u(t)\rangle = U(t)|u_i\rangle$ and  $\langle \tilde u(t) = \langle \tilde u_i| U^{-1}(t)$, where
\begin{eqnarray}
U(t)& = &\bigg(\cos\frac{\Omega t}{2} - i\sin\frac{\Omega t}{2}\,{\boldsymbol {\hat \Omega}}\cdot \boldsymbol \sigma\bigg) e^{-i\lambda_0 t/2}
\label{Eq31} \\
U^{-1}(t)& = &\bigg (\cos\frac{\Omega t}{2} + i\sin\frac{\Omega t}{2}\,{\boldsymbol {\hat \Omega}}\cdot \boldsymbol \sigma \bigg) e^{i\lambda_0 t/2}
\label{Eq31a}
\end{eqnarray}
$\boldsymbol {\hat \Omega}$ being the complex unit vector and $\Omega = (\boldsymbol {\Omega}\cdot\boldsymbol { \Omega})^{1/2}$.

Let $|u_i\rangle$ and $|u_f \rangle = |u(t)\rangle $ be initial and final states, respectively. Denoting the associated adjoint states by $\langle \tilde u_i|$ and $\langle \tilde u_f|$,  we compute the transition amplitude $|u_i\rangle \rightarrow |u_i\rangle$ and $|u_i\rangle \rightarrow |u_f\rangle$ as
\begin{eqnarray}
  T_{ii} &=&  \langle \tilde u_i |U(t)| u_i\rangle = \bigg( \cos\frac{\Omega t}{2} - i\sin\frac{\Omega t}{2}\;{ \mathbf n_i \cdot\boldsymbol {\hat \Omega}}\bigg) e^{-i\lambda_0t/2}  \label{T1}\\
   T_{fi} &=&  \langle \tilde u_f |U(t)| u_i\rangle = \bigg( \cos\theta_{fi} \cos\frac{\Omega t}{2} - i\sin\frac{\Omega t}{2}\;{\mathbf n_{fi}\cdot\boldsymbol {\hat \Omega}}\bigg) e^{i\lambda_0t/2}  \label{T1a}
\end{eqnarray}
where $ \cos\theta_{fi}= \langle \tilde u_f| u_i\rangle$, $\mathbf n_i = \langle \tilde u_i |\boldsymbol \sigma| u_i\rangle$ and  $\mathbf n_{fi} = \langle \tilde u_f |\boldsymbol \sigma| u_i\rangle$.

The computation of the time dependent Bloch vector yields
\begin{eqnarray}
\mathbf n(t) =  \cos{\Omega t}\, \mathbf n_i + \cos\chi(1-  \cos{\Omega t}){ \boldsymbol{\hat \Omega}} +  \sin{\Omega t}\,({\boldsymbol{\hat \Omega}} \times\mathbf n_i  ),
\label{Eq32}
\end{eqnarray}
where $\chi$ is the angle between the vectors $\mathbf n_i$ and $\boldsymbol{\hat \Omega}$, so that is $ \cos\chi = \mathbf n_i \cdot\boldsymbol{\hat \Omega}$.

Of the special interest is the case, when  $\Im (\boldsymbol {\Omega}\cdot\boldsymbol { \Omega})=0$. Denoting by $\Omega_0 =(| \boldsymbol {\Omega}\cdot\boldsymbol { \Omega}|)^{1/2}$, we obtain
\begin{eqnarray}
 \fl\mathbf n(t)   &=& \cos{\Omega_0 t}\, \mathbf n_i + \cos\chi_0(1-  \cos{\Omega_0 t})\frac{ \boldsymbol{\Omega}}{\Omega_0} +  \frac{\sin{\Omega_0 t}}{\Omega_0}({\boldsymbol{\hat \Omega}} \times\mathbf n_i  ),\; {\rm if} \; \boldsymbol {\Omega}\cdot\boldsymbol { \Omega} > 0 \label{Eq34} \\
  \fl\mathbf n(t)   &=& \cosh{\Omega_0 t}\, \mathbf n_i - \cos\chi_0(1-  \cosh{\Omega_0 t})\frac{ \boldsymbol{\Omega}}{\Omega_0} +  \frac{\sinh{\Omega_0 t}}{\Omega_0}({\boldsymbol{\hat \Omega}} \times\mathbf n_i  ) ,\; {\rm if} \; \boldsymbol {\Omega}\cdot\boldsymbol { \Omega} < 0
   \label{Eq34a}
\end{eqnarray}
where $\cos\chi_0= \mathbf n_i \cdot\boldsymbol{\Omega}/\Omega_0$.  At the exceptional point given by $\Omega =0$ and $\boldsymbol{ \Omega}=\boldsymbol{ \Omega_e}$, both regimes yield
\begin{eqnarray}
\mathbf n(t) = \mathbf n_i  - t(\mathbf n_i  \times\boldsymbol{ \Omega_e}) + \frac{t^2}{2}(\mathbf n_i  \cdot \boldsymbol{ \Omega_e}) \boldsymbol{\Omega_e}.
\label{Eq32a}
\end{eqnarray}

\begin{figure}[tbh]
%\begin{minipage}[]{16cm}
\begin{center}
%%\psfrag{r}{\huge $ \rho$}
\scalebox{0.325}{\includegraphics{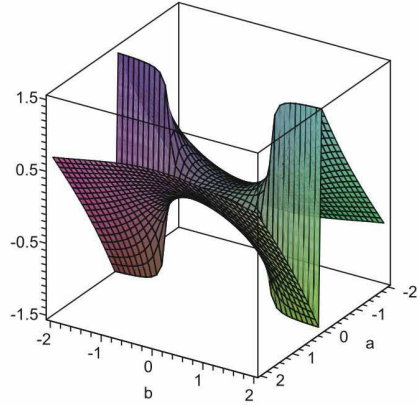}}
\end{center}
\caption{ The graphic of $\Re\gamma$ as function of $a=\Re(\mathbf n_i \cdot\boldsymbol{\Omega_e}) $ and $b=\Im(\mathbf n_i \cdot\boldsymbol{\Omega_e}) $ ($t=2$).
}\label{RGP3h}
%\end{minipage}
\end{figure}

Similar consideration of the transition amplitude yields
\begin{eqnarray}
\left.
\begin{array}{l}
  T_{ii} =   \bigg( \cos\frac{\Omega t}{2} - i\sin\frac{\Omega t}{2}\;{ \mathbf n_i \cdot\frac{ \boldsymbol{\Omega}}{\Omega_0}}\bigg) e^{-i\lambda_0t/2} \\
   T_{fi} =  \bigg( \cos\theta_{fi} \cos\frac{\Omega t}{2} - i\sin\frac{\Omega t}{2}\;{\mathbf n_{fi}\cdot \frac{ \boldsymbol{\Omega}}{\Omega_0}}\bigg) e^{i\lambda_0t/2}
\end{array}
\right \} \;\rm if \;  \boldsymbol {\Omega}\cdot\boldsymbol { \Omega}> 0 \label{T2}
\end{eqnarray}
and
\begin{eqnarray}
\left.
\begin{array}{l}
  T_{ii} = \bigg( \cosh\frac{\Omega t}{2} - i\sinh\frac{\Omega t}{2}\;{ \mathbf n_i \cdot \frac{ \boldsymbol{\Omega}}{\Omega_0}}\bigg) e^{-i\lambda_0t/2} \\
   T_{fi} = \bigg( \cos\theta_{fi} \cosh\frac{\Omega t}{2} - i\sinh\frac{\Omega t}{2}\;{\mathbf n_{fi}\cdot \frac{ \boldsymbol{\Omega}}{\Omega_0}}\bigg) e^{i\lambda_0t/2}
\end{array}
\right \} \;\rm if \;  \boldsymbol {\Omega}\cdot\boldsymbol { \Omega} < 0 \label{T2a}
\end{eqnarray}
Thus, if $ \boldsymbol {\Omega}\cdot\boldsymbol { \Omega} > 0$ we obtain {\em coherent} evolution of the quantum-mechanical system, and, if $ \boldsymbol {\Omega}\cdot\boldsymbol { \Omega} < 0$ we have {\em incoherent} one. At the exceptional point both regimes yield
\begin{eqnarray}
  T_{ii} = \bigg( 1 - i\frac{t}{2}\;{ \mathbf n_i \cdot{ \boldsymbol{\Omega_e}}}\bigg) e^{-i\lambda_0t/2} \label{T3a}\\
   T_{fi} = \bigg( \cos\theta_{fi}  - i\frac{t}{2}\;{\mathbf n_{fi}\cdot { \boldsymbol{\Omega_e}}}\bigg) e^{i\lambda_0t/2}
 \label{T3b}
\end{eqnarray}
\begin{figure}[tbh]
%\begin{minipage}[]{16cm}
%\begin{center}
%%\psfrag{r}{\huge $ \rho$}
\scalebox{0.325}{\includegraphics{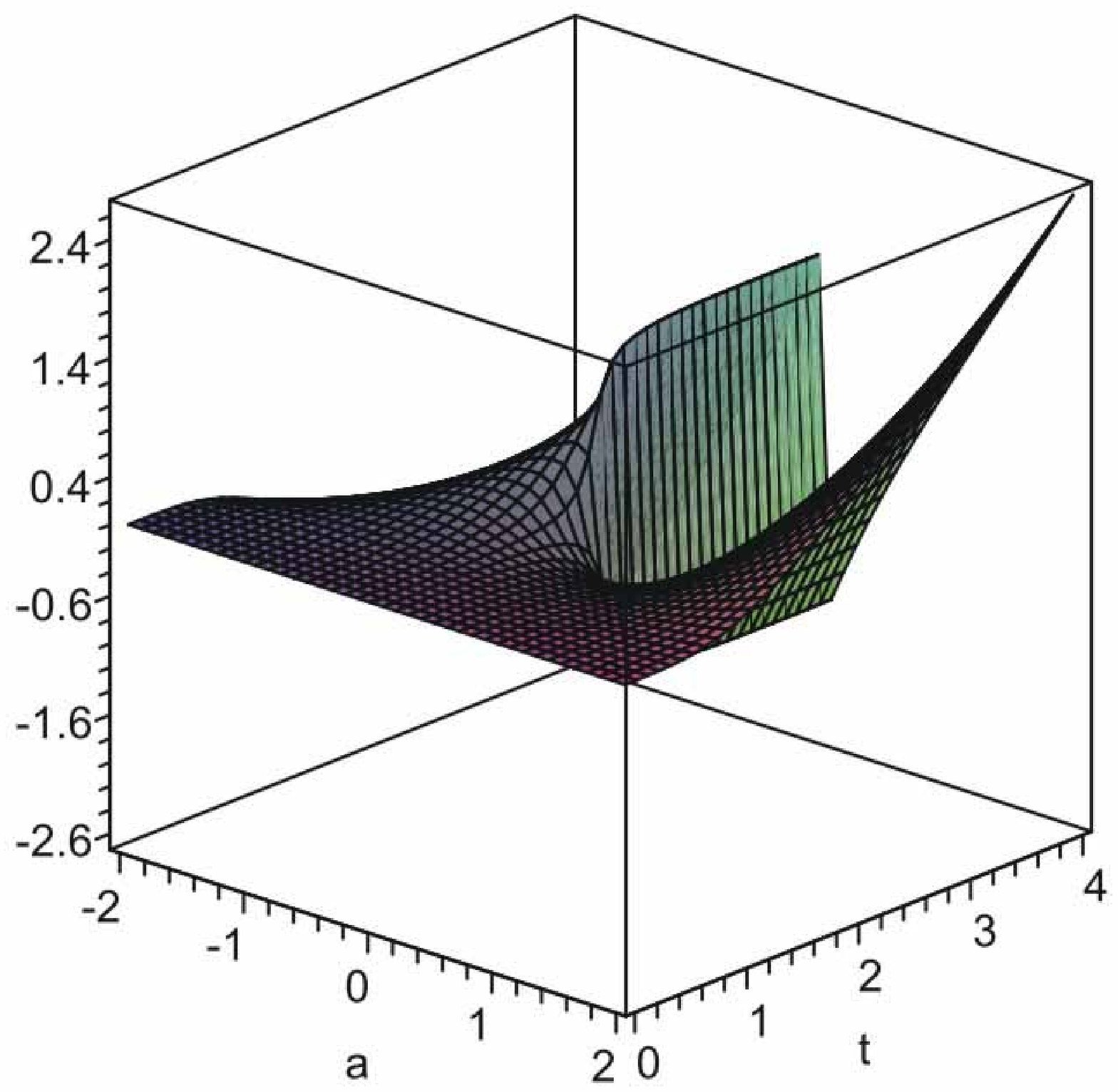}}
%\hspace{1cm}
\scalebox{0.325}{\includegraphics{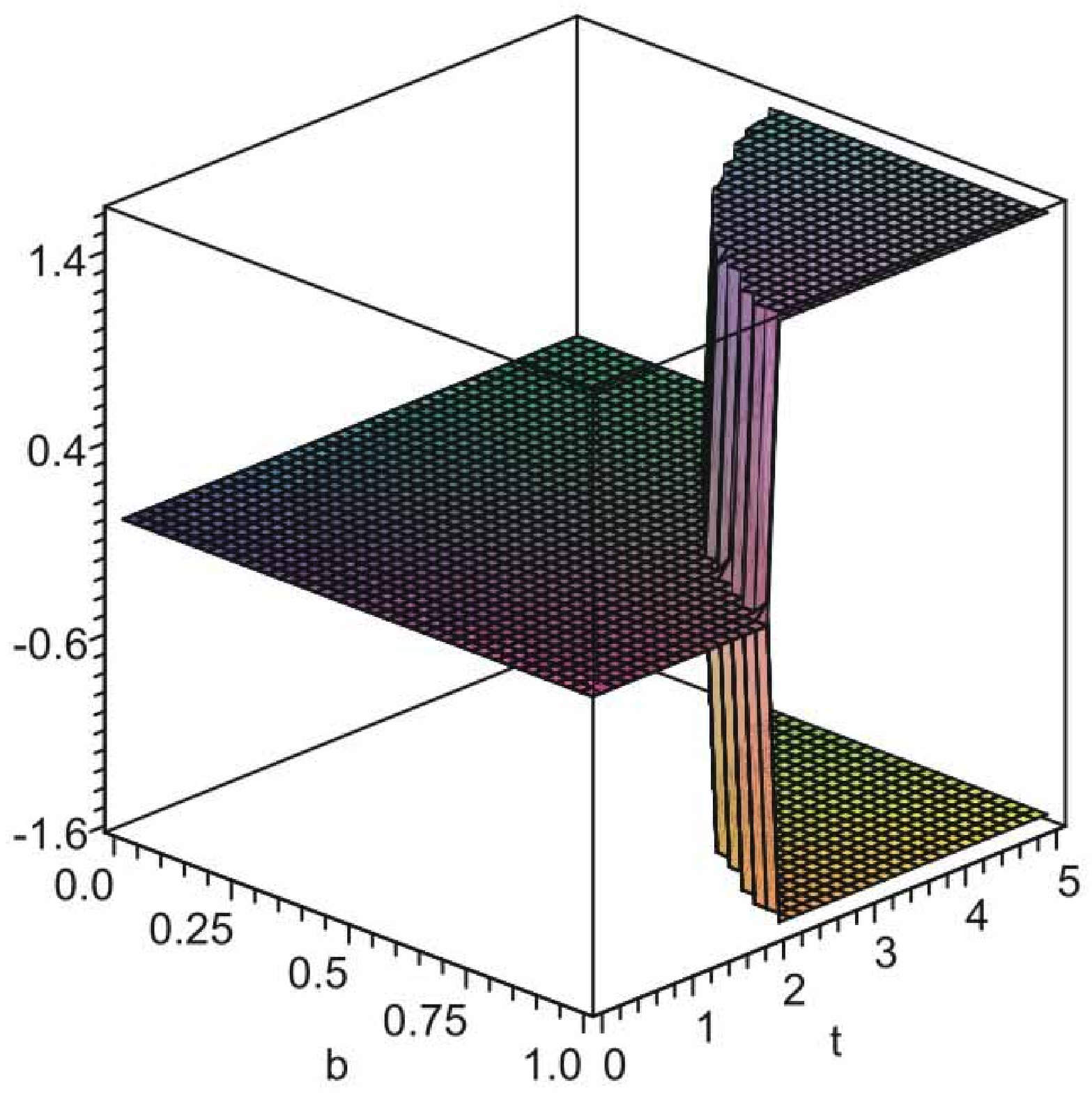}}
%\end{center}
\caption{Left panel: real part of geometric phase $\Re\gamma$ vs. $a=\Re(\mathbf n_i \cdot\boldsymbol{\Omega_e}) $ and $t$ ($\Re(\mathbf n_i \cdot\boldsymbol{\Omega_e}) = 1$). Right panel: $\Re\gamma$ vs. $b=\Im(\mathbf n_i \cdot\boldsymbol{\Omega_e}) $ and $t$  ($\Re(\mathbf n_i \cdot\boldsymbol{\Omega_e}) =0$).
}\label{RGP3f}
%\end{minipage}
\end{figure}
The complex geometric phase can be derived from the formula (\ref{GP5a}), or equivalently, using (\ref{Eq28b}).  Employing (\ref{GP5a}) we obtain the time dependent geometric phase as
\begin{eqnarray}
\gamma (t) =\frac{\Omega t}{2}  \cos\chi +\frac{i}{2} \ln\frac{1 +i \cos\chi\, \tan \frac{\Omega t}{2} }{1 -i \cos\chi\, \tan \frac{\Omega t}{2}}.
\label{GP6}
\end{eqnarray}
In the vicinity of the exceptional point we have
\begin{eqnarray}
\gamma(t) =\mathbf n_i \cdot\boldsymbol{\Omega_e}\, \frac{ t}{2} +\frac{i}{2} \ln\frac{1+ i\mathbf n_i \cdot\boldsymbol{\Omega_e}  \,\frac{ t}{2}}{1- i \mathbf n_i \cdot\boldsymbol{\Omega_e} \, \frac{ t}{2}} + {\cal O }(\Omega^2).
\label{GP6a}
\end{eqnarray}
It follows from here that the real part of the geometric phase $\Re \gamma(t)$ behaves like a step-function at the point $t_0 = 2/{|\Im(\mathbf n_i \cdot\boldsymbol{\Omega_e})|}$ (see Figs. \ref{RGP3f} \ref{RGP3h}). In particular,  we obtain
\begin{eqnarray}
\fl
\Re\gamma(t_0) = \lim_{\Re(\mathbf n_i \cdot\boldsymbol{\Omega_e}) \rightarrow 0}\,\lim_{t \rightarrow t_0}\Re\gamma(t) =\left \{
\begin{array}{rll}
  0, &\Re(\mathbf n_i \cdot\boldsymbol{\Omega_e}) \rightarrow 0, & t \rightarrow t_0 -0\\
  -\pi/2, &\Re(\mathbf n_i \cdot\boldsymbol{\Omega_e}) \rightarrow +0, & t \rightarrow t_0 + 0\\
   \pi/2,& \Re(\mathbf n_i \cdot\boldsymbol{\Omega_e}) \rightarrow -0, &t \rightarrow t_0 + 0
    \end{array}
\right .
\label{GP7b}
\end{eqnarray}

\subsection{Two-level atom driven by periodic electromagnetic field}

As an illustrative example we consider a two-level dissipative system driven by periodic electromagnetic field $\mathbf E(t)= \Re (\boldsymbol{\mathcal E}(t)\exp(i\nu t))$. In the rotating wave approximation, after removing the explicit time dependence of the Hamiltonian and the average effect of the decay terms, the Schr\"odinger equation reads \cite{GW,LSS}
\begin{eqnarray}\label{Sch1e}
i\frac{\partial}{\partial t}\left(
  \begin{array}{c}
    u_1 \\
   u_2 \\
  \end{array}
\right) =
\frac{1}{2}\left(
  \begin{array}{cc}
   -i\lambda+ \Delta  - i\delta & 2V^\ast \\
   2 V & -i\lambda - \Delta + i\delta \\
  \end{array}
  \right)
\left(
  \begin{array}{c}
    u_1 \\
    u_2 \\
  \end{array}
\right)
\end{eqnarray}
where $\lambda = (\gamma_a +\gamma_b)/2$ with $\gamma_a, \;\gamma_b$ being decay rates for upper and lower levels,
respectively, $\Delta= \omega_0 - \nu$, $\omega_0= (E_a-E_b)$, $\delta =(\gamma_a- \gamma_b)/2$, $V=\boldsymbol\mu \cdot \boldsymbol{\mathcal E}$ and $\boldsymbol \mu$ is the electric dipole moment.

The choice $\boldsymbol{\mathcal E}(t) = \boldsymbol{\mathcal E}_0 \exp (i\omega t) $ yields $V(t)= V_0\exp (i\omega t)$, where $V_0 = \boldsymbol\mu \cdot \boldsymbol{\mathcal E}_0$, and we assume further $V_0 >0$. The solution of Eq. (\ref{Sch1e}) with this choice of $\boldsymbol{\mathcal E}$ is well known and can be written as
\begin{eqnarray}
\label{Sol1}
|u(t)\rangle= C_1(t)e^{-i(\omega - i\lambda)t/2}|u_{\uparrow}\rangle
+  C_2(t)e^{i(\omega + i\lambda)t/2}|u_{\downarrow}\rangle
\end{eqnarray}
where  and $|u_{\uparrow}\rangle= \left(
  \scriptsize\begin{array}{c}
  1 \\
 0 \\
  \end{array}\right)$,
$|u_{\downarrow}\rangle= \left(
  \scriptsize\begin{array}{c}
  0 \\
 1 \\
  \end{array}\right)$ denote the up/down states, respectively.
In addition, $|C (t)\rangle$  satisfies the Schr\"odinger equation
\begin{eqnarray}\label{Sch1a}
i\frac{\partial |C \rangle }{\partial t}= H_r  |C \rangle
\end{eqnarray}
written in the co-rotating reference frame, where the Hamiltonian of the system takes the form
\begin{eqnarray}\label{HR}
H_r =
\frac{1}{2}\left(
  \begin{array}{cc}
    \Delta -\omega  - i\delta & 2V_0\\
  2 V_0 & - \Delta + \omega + i\delta \\
  \end{array}
  \right)
\end{eqnarray}
and we find
\begin{eqnarray}\label{Sol2}
\fl \left(
  \begin{array}{c}
  C_1(t) \\
  C_2(t) \\
  \end{array}
\right) =
\left(
  \begin{array}{cc}
    \cos(\Omega t/2)-i\cos\chi\sin(\Omega t/2) &-i\sin\chi\sin(\Omega t/2)\\
    -i\sin\chi\sin(\Omega t/2) & \cos(\Omega t/2)+i\cos\chi\sin(\Omega t/2) \\
  \end{array}
  \right)
\left(
  \begin{array}{c}
    C_1(0) \\
    C_2(0) \\
  \end{array}
\right), \nonumber\\
\end{eqnarray}
where $\cos\chi = (\Delta - \omega -i\delta)/\Omega$, $\Omega=(\rho^2 + (\Delta - \omega -i\delta)^2)^{1/2}$, $\rho=2V_0$.

Without loss of generality, we can further confine out attention to the case e $\varphi =0$, that implies $V_0 = |V_0|$. Passing on to the Bloch vector $\mathbf n(t) = \langle \tilde u(t)|\boldsymbol\sigma|u(t)\rangle$ we obtain
\begin{eqnarray}
\label{B2}
\fl
\mathbf n(t) = \left(\begin{array}{ccc}
\cos\omega t& -\sin\omega t & 0\\
\sin\omega t & \cos\omega t & 0 \\
0 & 0 & 1 \\
\end{array}\right)
\left(\begin{array}{ccc}
\sin^2\chi+\cos^2\chi\cos\Omega t & -\cos\chi\sin\Omega t& \frac{1}{2} \sin2\chi(1- \cos\Omega t)\\
\cos\chi\sin\Omega t &  \cos\Omega t &  -\sin\chi\sin\Omega t\\
\frac{1}{2} \sin2\chi(1- \cos\Omega t)&\sin\chi\sin\Omega t &\cos^2\chi+\sin^2\chi\cos\Omega t\\
\end{array}\right)\mathbf n(0) \nonumber\\
\end{eqnarray}
Finally, one can show that  $\mathbf n(t)$ satisfies the following equation
\begin{eqnarray}\label{B1a}
    d \mathbf n/dt= \mathbf\Omega^\prime(t) \times \mathbf n
\end{eqnarray}
where $\mathbf\Omega^\prime =(\rho\cos\omega t,\rho\sin\omega t, \Delta - i\delta)$.

\subsubsection{Cyclic evolution.} Let $\mathbf n(t +T)=\mathbf n(t)$ be he Bloch vector yielding a cyclic evolution of system over the complex sphere $S^2_c$ with the period  $T= 2\pi/\omega$. Starting with its definition $\mathbf n = \langle \tilde u(t)|\boldsymbol \sigma |u(t)\rangle$, where  $|u(t)\rangle$ and $\langle \tilde u(t)|$  satisfy the Schr\"odinger equation (\ref{Sch1e}) and is its adjoint equation, respectively, we find that the solution
\begin{eqnarray}
\label{Sol3}
|u_+(t)\rangle= \cos\frac{\chi}{2}\;e^{-i(\omega + \Omega- i\lambda)t/2}|u_{\uparrow}\rangle
+ \sin\frac{\chi}{2} \, e^{i(\omega - \Omega + i\lambda)t/2}|u_{\downarrow}\rangle\\
\langle \tilde u_+(t)|= \cos\frac{\chi}{2}\;e^{i(\omega + \Omega- i\lambda)t/2}\langle u_{\uparrow}|
+ \sin\frac{\chi}{2} \, e^{-i(\omega - \Omega + i\lambda)t/2}|\langle u_{\downarrow}|
\end{eqnarray}
yields
\begin{equation}
\mathbf n_+ = (\sin \chi \cos\omega t, \sin \chi \sin\omega t, \cos\chi ).
\end{equation}
Note that $\mathbf n_+(t)$ can be obtained as the periodic solution of the Bloch equation with  $\boldsymbol \Omega = \Omega(\sin \chi, 0,\cos \chi)$. The other periodic solutionis is given by
\begin{equation}
\mathbf n_{-} =- (\sin \chi \cos\omega t, \sin \chi \sin\omega t, \cos\chi ),
\end{equation}
and the related solution of the Schr\"odinger equation reads:
\begin{eqnarray}
\label{Sol3a}
|u_{-}(t)\rangle= -\sin\frac{\chi}{2}\;e^{-i(\omega + \Omega- i\lambda)t/2}|u_{\uparrow}\rangle
+ \cos\frac{\chi}{2} \, e^{i(\omega - \Omega + i\lambda)t/2}|u_{\downarrow}\rangle\\
\langle \tilde u_{-}(t)|= -\sin\frac{\chi}{2}\;e^{i(\omega + \Omega- i\lambda)t/2}\langle u_{\uparrow}|
+ \cos\frac{\chi}{2} \, e^{-i(\omega - \Omega + i\lambda)t/2}|\langle u_{\downarrow}|
\end{eqnarray}

The geometric phase derived from (\ref{Eq27g}) is given by
\begin{eqnarray}\label{36}
    \gamma_{\pm}= -\pi\bigg(1 \mp \frac{\Delta - \omega -i\delta}{\sqrt{\rho^2 + (\Delta - \omega -i\delta)^2}}\bigg)
\end{eqnarray}
In the adiabatic limit, $|\omega/(\Delta - i\delta)|\ll 1$, the complex Aharonov-Anandan phase $\gamma = \gamma_{-} + 2\pi$ reduces to the complex Berry phase $\gamma_{ad}$ obtained in \cite{GW}:
\begin{equation}\label{Eq30g}
\gamma_{ad}= \pi\bigg( 1- \frac{\Delta - i\delta}{\sqrt{(2V_0)^2+ (\Delta -
i\delta)^2}}\bigg).
\end{equation}
where $2V_0 = \rho$.

\begin{figure}[tbh]
%\begin{minipage}[]{16cm}
%\begin{center}
%\psfrag{r}{\huge $ \rho$}
\scalebox{0.325}{\includegraphics{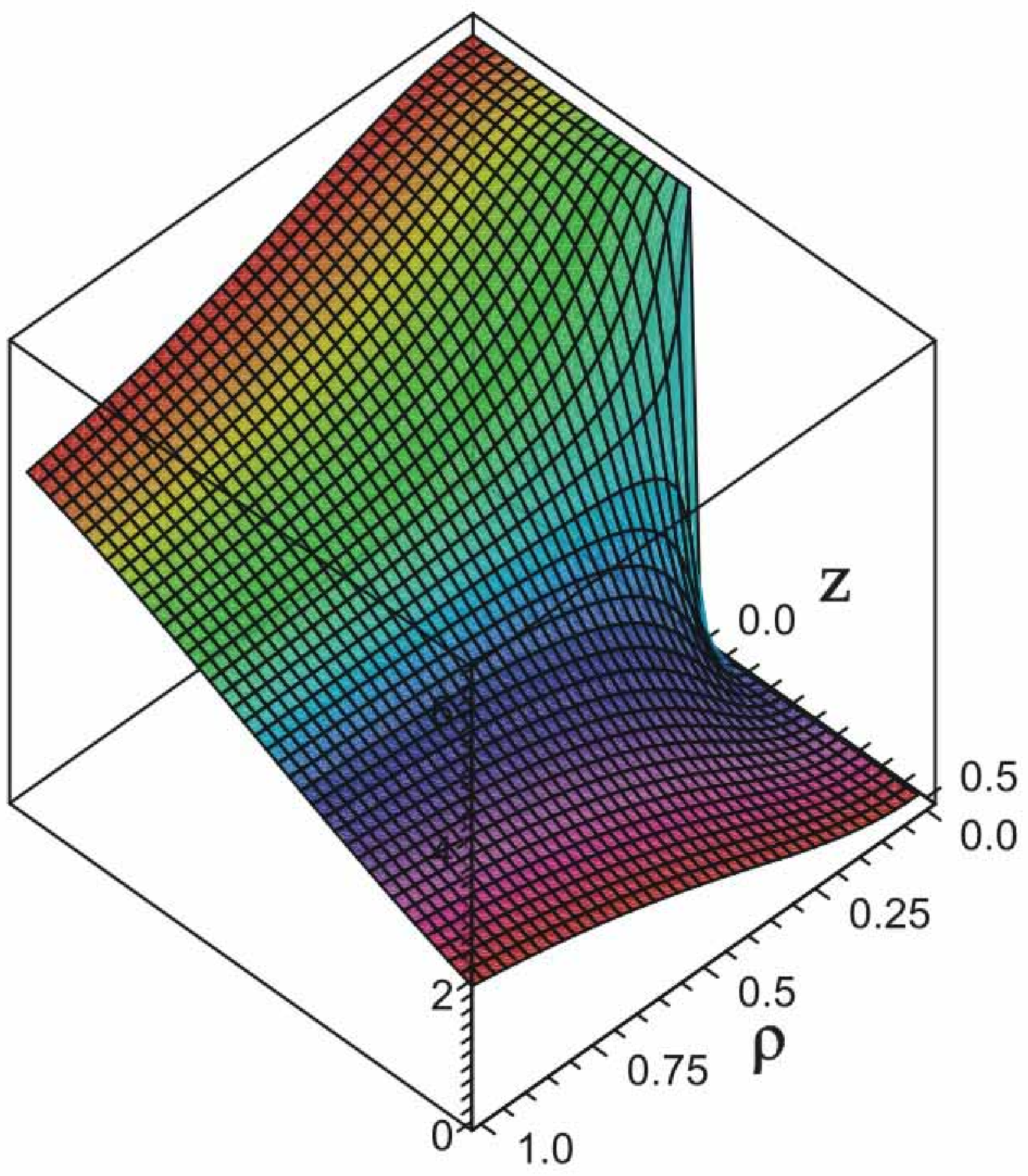}}
\hspace{1cm}
\scalebox{0.325}{\includegraphics{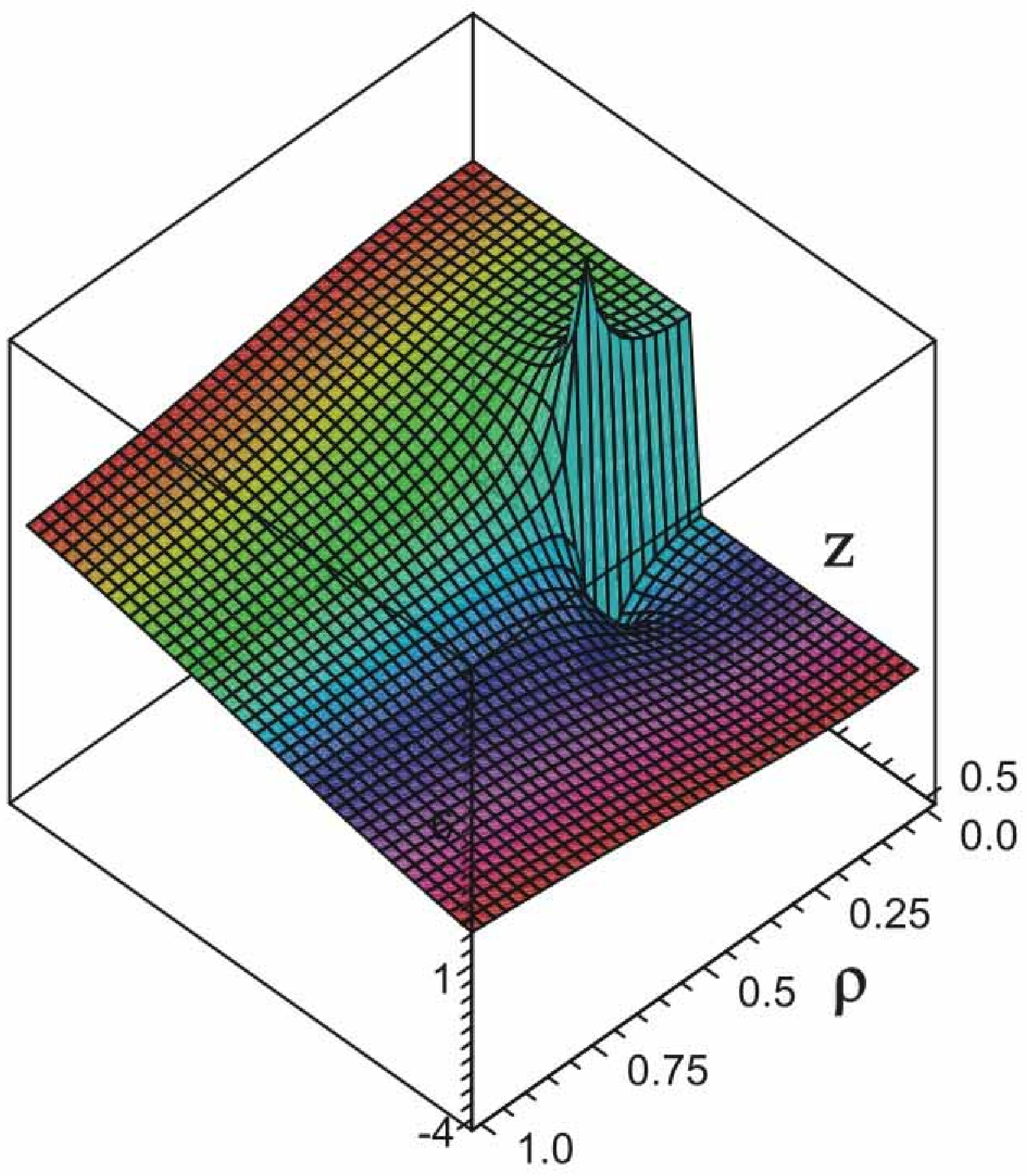}}
%\end{center}
\caption{Left panel: real part of geometric phase, $\Re\gamma$ vs $\rho$ and $z= \Delta - \omega$, in the vicinity of the diabolic point given by $\rho=z=0$ ($\delta =0$). Right panel: $\Re\gamma$ vs $\rho$ and $z$ nearby the exceptional point defined by $\rho=\delta$ and $ z=0$ ($\delta =0.25$).
}\label{RP1c}
%\end{minipage}
\end{figure}

In what follows we consider the behavior of the geometric phase $\gamma$ near the critical points, starting with the diabolic point. In this case $\Im \gamma =0$ and we have
\begin{equation}\label{Q4}
\gamma= \pi\bigg(1 - \frac{\Delta - \omega }{\sqrt{\rho^2 + (\Delta - \omega )^2}}\bigg).
\end{equation}
This yields
\begin{eqnarray}\label{GP1a}
 \gamma= \bigg \{
\begin{array}{l}
 0, \;{\rm for}\; \rho =0, \; \Delta - \omega  >0 \\
 2\pi, \; {\rm for} \; \rho =0, \; \Delta - \omega  <0
\end{array}
\end{eqnarray}
It follows that the geometric phase behaves as the step-function near
the diabolic point, and at the diabolic point one has the discontinuity of $\gamma$ with the gap of $2\pi$ (Fig. \ref{RP1c}).

Once return to (\ref{36}), we find that near the exceptional point, and for $\Delta =\omega$, real part of the geometric phase is given by
\begin{eqnarray}
 \Re \gamma= \Bigg \{
\begin{array}{l}
 \pi, \;{\rm if}\; \rho > \delta  \\
  \pi \Big(1\pm \displaystyle\frac{\delta}{\sqrt{\delta^2 - \rho^2 }}\Big), \; {\rm if} \;
  \rho < \delta
\end{array}
\end{eqnarray}
where the upper/lower sign corresponds to $\Delta - \omega  \rightarrow \pm 0$. Similar consideration of imaginary part of the geometric phase yields
\begin{eqnarray}\label{Q5}
 \Im \gamma= \Bigg \{
\begin{array}{l}
 0, \;{\rm if}\; \rho < \delta \\
   \displaystyle\frac{\pi\delta}{\sqrt{ \rho^2 - \delta^2 }}, \; {\rm if} \;
  \rho >\delta
\end{array}
\end{eqnarray}
As can be observed in Fig. \ref{RP1c},  the geometric phase has an infinite gap at the exceptional point. This is due to the coalescence of eigenvectors at the exceptional point.

\subsubsection{Non-cyclic evolution.}

Let us consider the case in which the initial state is $|u(0)\rangle =|u_\uparrow\rangle$, which corresponds to the north pole of the Bloch sphere $S^2_c$, and, hence, $\mathbf n(0)=(0,0,1)$. For non-cyclic evolution and the initial state chosen as $\mathbf n_i= (0,0,1)$, the explicit form of the time dependent solution of Eq.(\ref{B1}) is given by
\begin{eqnarray}
&\mathbf n(t) = \left(\begin{array}{l}
\sin\chi \cos\chi(1-\cos\Omega t)\cos\omega t + \sin\chi \sin\Omega t \sin\omega t\\
\sin\chi \cos\chi(1-\cos\Omega t)\sin\omega t-\sin\chi \sin\Omega t\cos\omega t \\
\cos^2\chi + \sin^2 \chi \cos\Omega t
\end{array}\right).
 \label{eq1a}
\end{eqnarray}

\begin{figure}[tbh]
%\begin{minipage}[]{16cm}
%\begin{center}
%\psfrag{r}{\huge $ \rho$}
\scalebox{0.325}{\includegraphics{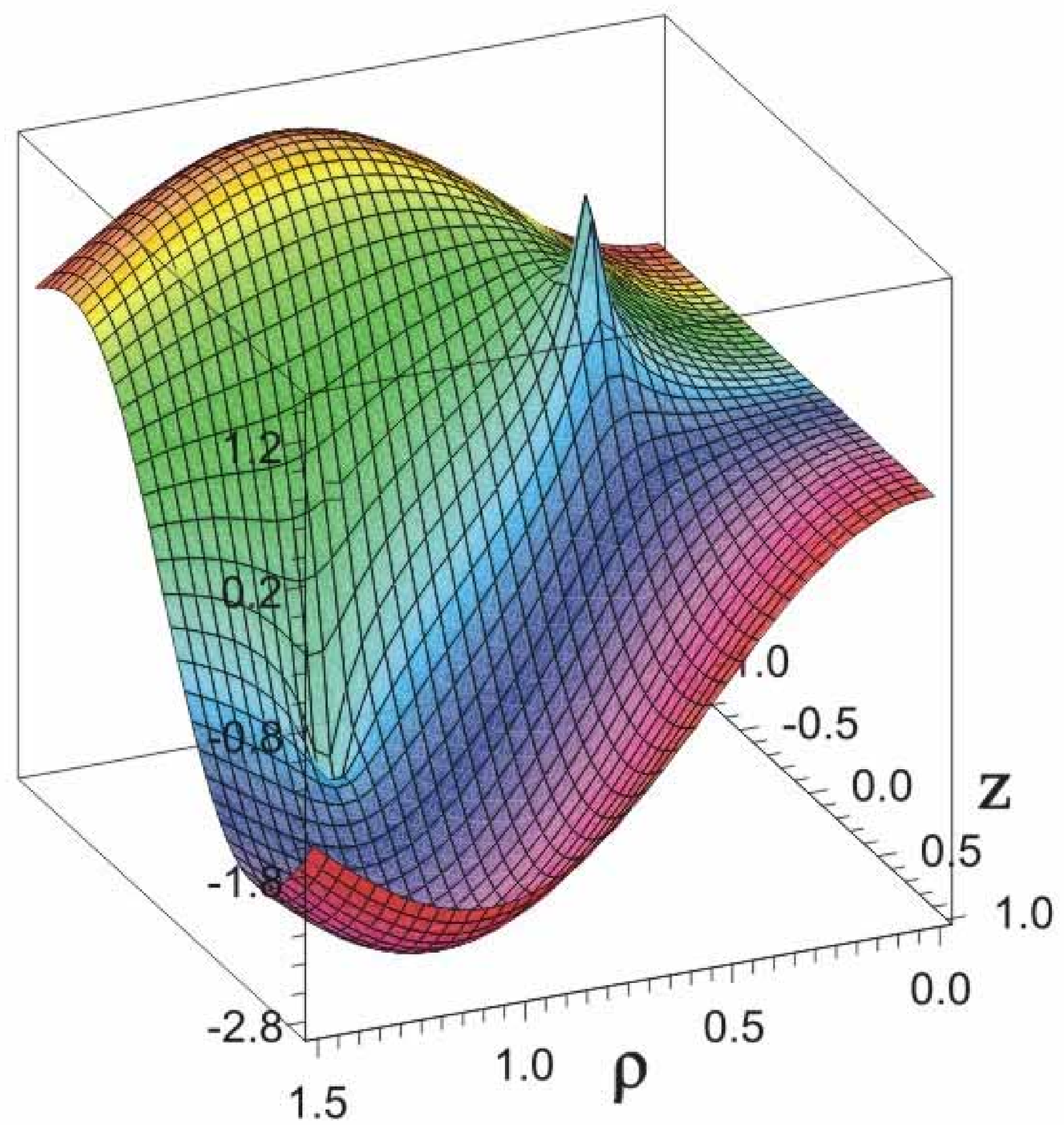}}
\hspace{1cm}
\scalebox{0.325}{\includegraphics{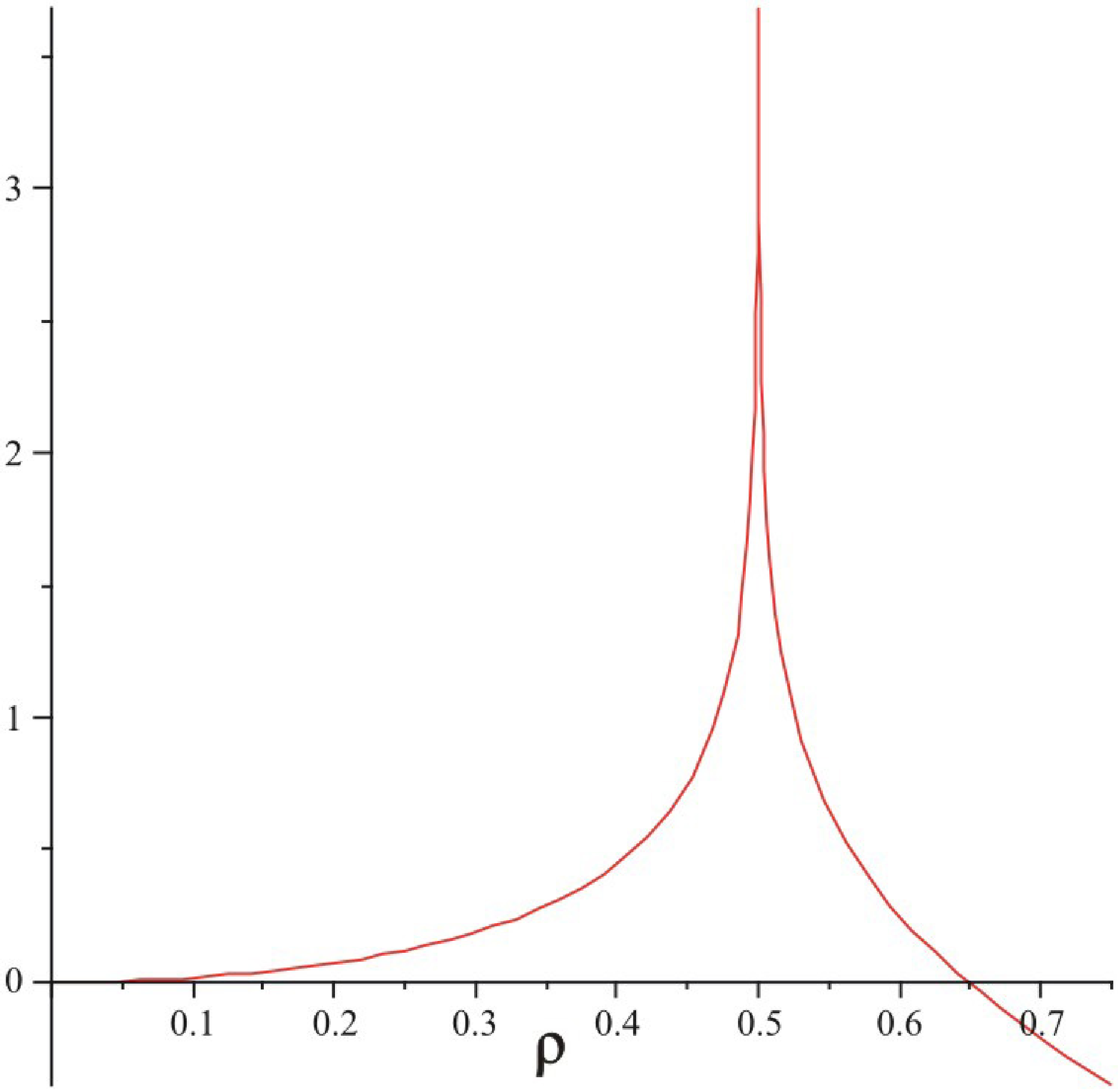}}
%\end{center}
\caption{Left panel: imaginary part of the geometric phase, $\Im\gamma$ as function of $\rho$ and $ z=\Delta - \omega$ ($t = 2/\delta,\; \delta =0.5, \;\omega =1$). Right panel:  $\Im\gamma$ vs. $\rho$ ($z=0, t = 2/\delta,\; \delta =0.5, \;\omega =1$). The divergence of $\Im\gamma$ can be observed at the exceptional point ($\rho=\delta =0.5, \; z=0$).
}
\label{EP1g}
%\end{minipage}
\end{figure}

\begin{figure}[tbh]
%\begin{minipage}[]{16cm}
%\begin{center}
%\psfrag{r}{\huge $ \rho$}
\scalebox{0.25}{\includegraphics{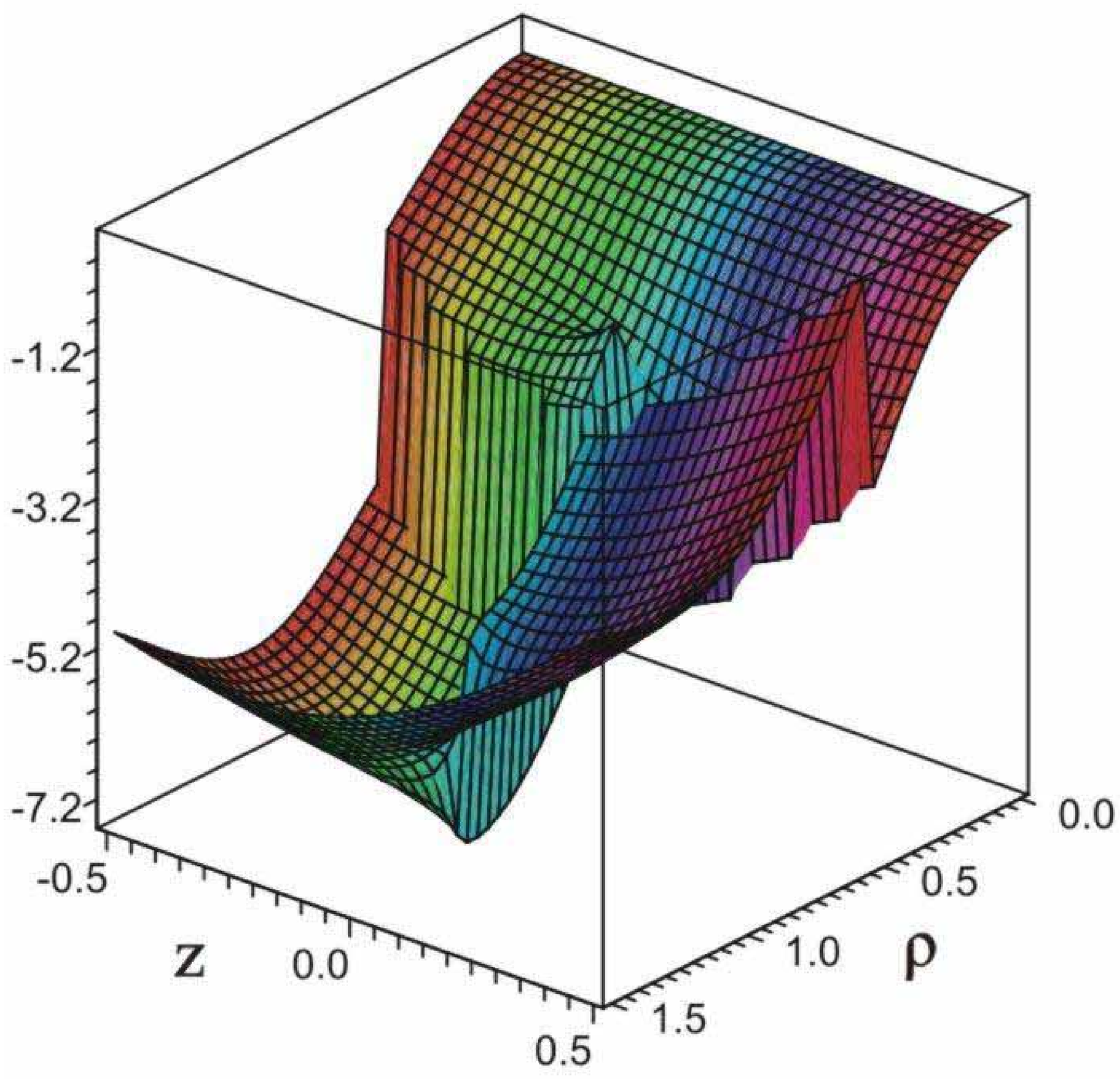}}
%\hspace{1cm}
\scalebox{0.375}{\includegraphics{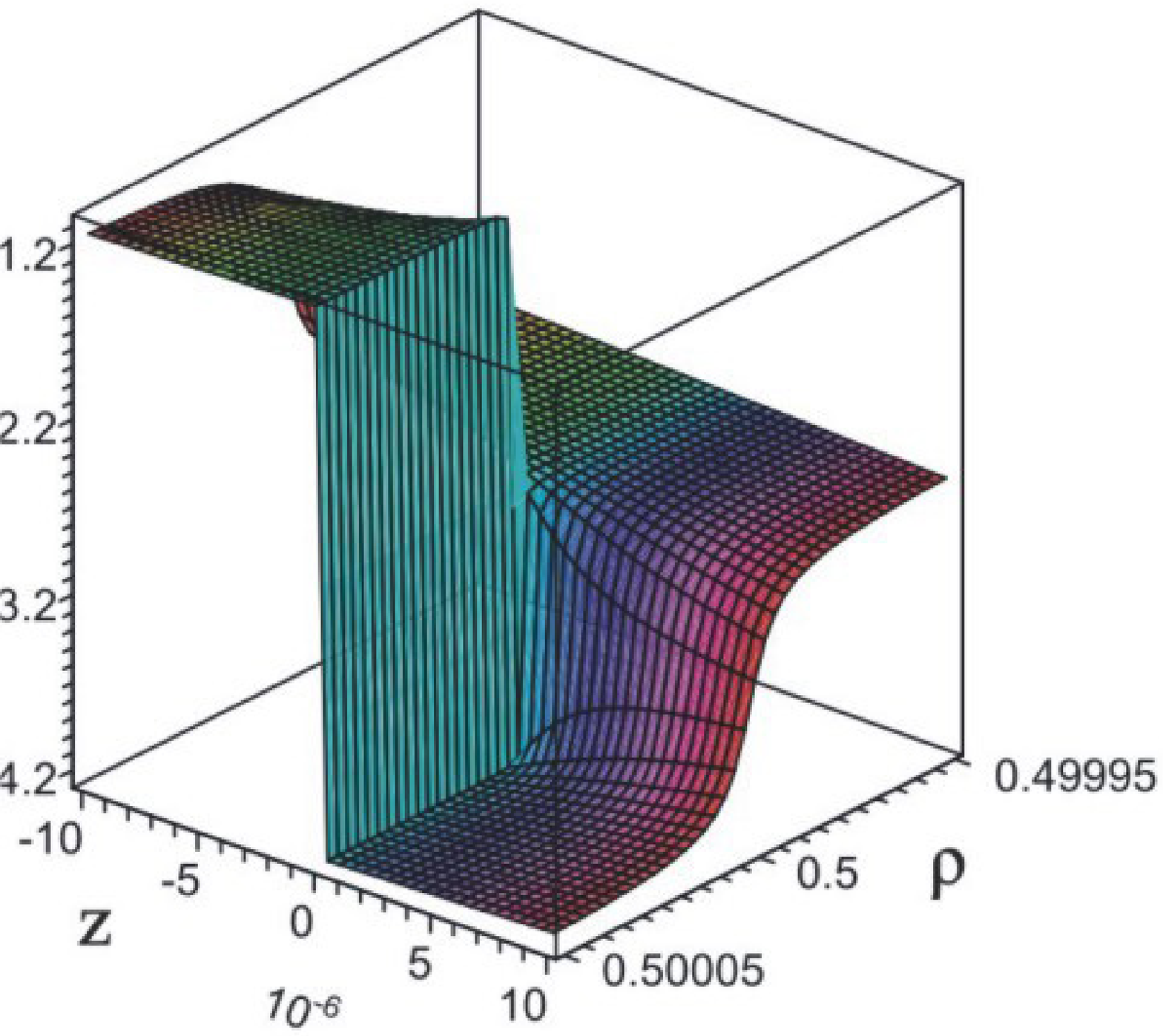}}
%\end{center}
\caption{Left panel: real part of geometric phase, $\Re\gamma$ vs $\rho$ and $z = \Delta - \omega$ ($t= 2/\delta = 0.5$, $\delta =0.5,\; \omega =1$). Right panel: the same graphic nearby the exceptional point defined by $\rho=\delta$ and $ z=0$.
}\label{RP1g}
%\end{minipage}
\end{figure}

The geometric phase derived from (\ref{Eq28b}) is given by
\begin{eqnarray}
\label{GP8}
\gamma = \frac{\Omega t}{2}\cos\chi -\frac{\omega \sin^2\chi}{2\Omega}(\Omega t - \sin \Omega t) + \frac{i}{2} \ln\frac{1 + i \cos\chi\tan\frac{\Omega t}
{2}}{1 - i \cos\chi\tan\frac{\Omega t}{2}}
\end{eqnarray}
where $\Omega= (\rho^2+ (\Delta-\omega -i\delta)^2)^{1/2}$. As depicted in Figs. \ref{EP1g} and \ref{RP1g}, the geometric phase has a singular behavior in vicinity of the exceptional point. At the latter, the imaginary part of the geometric phase shows the divergence and its real part behaves as step-like function.

In the vicinity of the degeneracy point defined by $\Omega =0$ we have
\begin{eqnarray}
\label{GP10}
\gamma =  \frac{Z t}{2} -\frac{\omega \rho^2t^3}{6} + \frac{i}{2}{\ln\frac{1+ i\frac{Zt}{2}(1+ \frac{1}{3}(\frac{\Omega t}{2})^2)}{1- i\frac{Zt}{2}(1+ \frac{1}{3}(\frac{\Omega t}{2})^2)}}  + {\cal O} (\Omega^4),
\end{eqnarray}
where $Z= \Delta - \omega -i\delta$, and we set $k=1$. If $t \neq 2/\delta$, it follows that at the exceptional point, defined by $Z=-i\delta$ and $\rho = \delta$, the geometric phase is given by
\begin{eqnarray}
\label{GP11}
\gamma =  -\frac{\omega \delta^2 t^3}{6}  -i \frac{\delta t}{2} + \frac{i}{2}{\ln\frac{1+\frac{\delta t}{2}}{1- \frac{\delta t}{2}}}
\end{eqnarray}

\begin{figure}[tbh]
%\begin{minipage}[]{16cm}
%\begin{center}
%\psfrag{\rho}{\Huge $ \rho$}
%\psfrag{z}{\huge $ \mathbf z$}
\scalebox{0.325}{\includegraphics{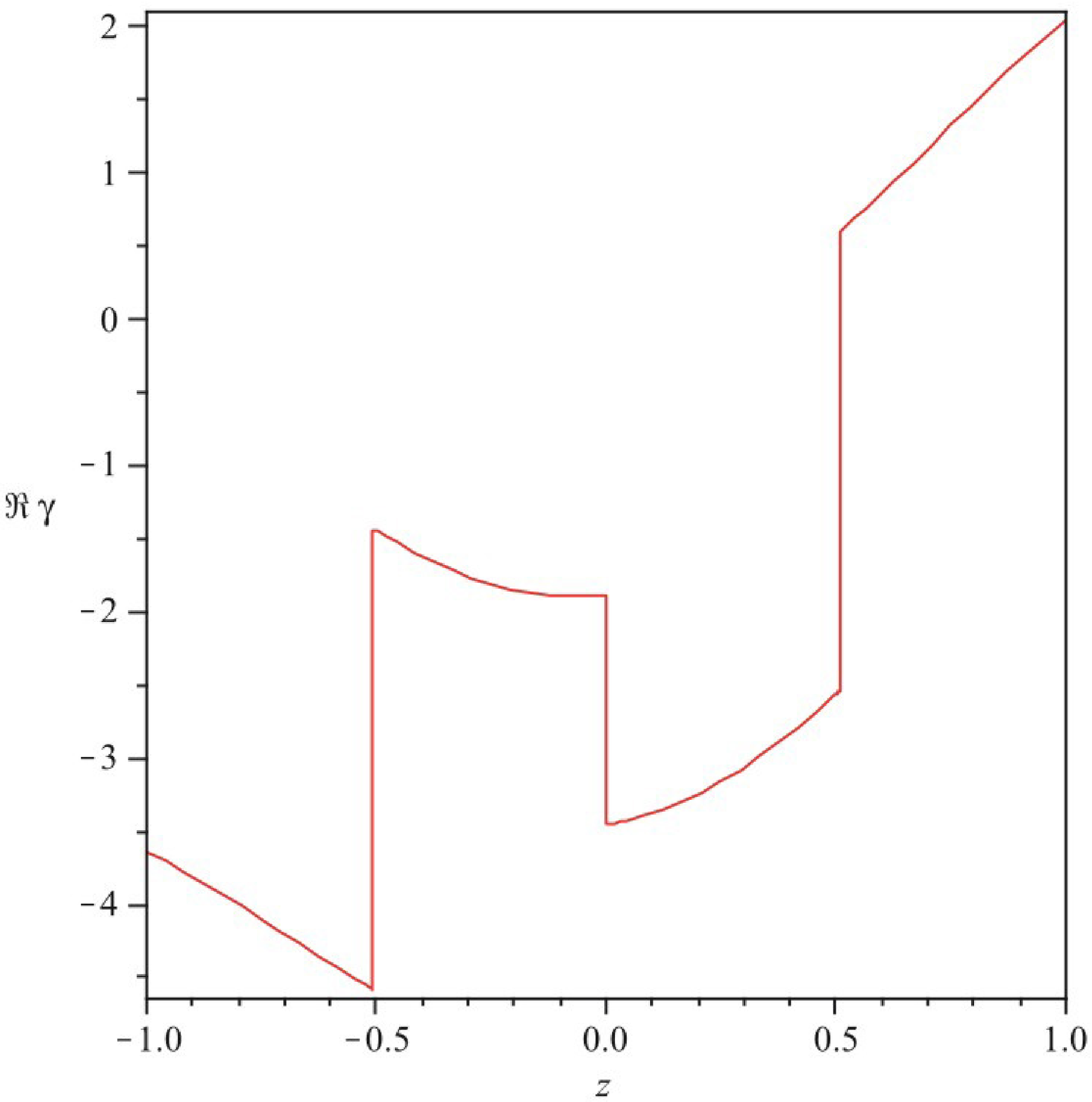}}
\hspace{1cm}
\scalebox{0.325}{\includegraphics{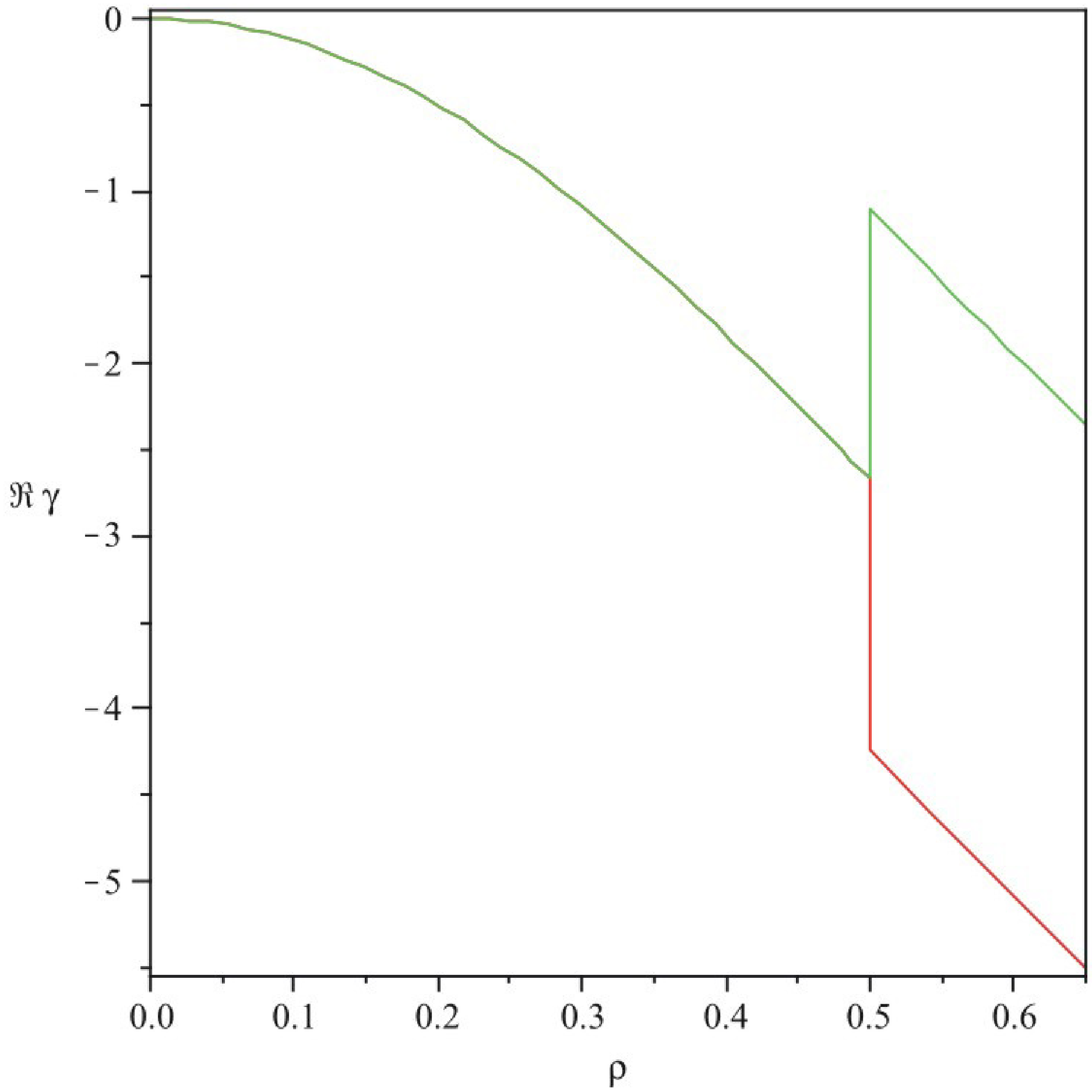}}
%\end{center}
\caption{Left panel: graphic of $\Re\gamma$ vs $z= \Delta - \omega$ depicted for $t= 2/\delta$ and $ \rho=\delta$ ($\delta =0.5,\, \omega =1$). At the exceptional point, defined by $\rho=\delta$ and $ z=0$, the jump discontinuity is given by $\Delta\Re\gamma = \pm\pi/4$. Right panel:  $\Re\gamma$ vs. $\rho$ ($t= 2/\delta$, $z= 0$, $\delta =0.5,\, \omega =1$). At the exceptional point $\Delta\Re\gamma = \pm\pi/2$.
}\label{RP1h}
%\end{minipage}
\end{figure}

The case of $t=2/\delta$ requires more careful analysis. Assuming $t=2/\delta$ and inserting $\rho=\delta$ into Eq.(\ref{GP10}) we obtain
\begin{eqnarray}
\label{GP12}
\gamma =  \pm \frac{\pi}{4}-\frac{4\omega }{3\delta}  + \frac{\Delta - \omega}{\delta } - \frac{i}{2}{\ln\frac{e^2|\Delta - \omega|}{2\delta +i(\Delta - \omega)}} + {\cal O} (\Omega^4)
\end{eqnarray}
where the upper/lower sign corresponds to $\rho - \delta \rightarrow \pm 0$. At the exceptional point we have (Fig. \ref{RP1h}, left panel)
 \begin{eqnarray}
\label{GP13}
\Re\gamma =  \pm \frac{\pi}{4} -\frac{4\omega }{3\delta} ,
\end{eqnarray}
In a similar way, assuming $\Delta- \omega =0$,  we obtain
\begin{eqnarray}
\label{GP12a}
\gamma =  \pm \frac{\pi}{2}-\frac{4\omega \rho^2}{3\delta^3}  - \frac{i}{2}{\ln\frac{|\rho^2 -\delta^2| e^2}{ \rho^2 + 5\delta^2}} + {\cal O} (\Omega^4)
\end{eqnarray}
where the upper/lower sign corresponds to $\Delta -\omega \rightarrow \pm 0$. At the exceptional point we have (Fig. \ref{RP1h}, right panel)
 \begin{eqnarray}
\label{GP13a}
\Re\gamma = \pm \frac{\pi}{2} -\frac{4\omega }{3\delta} ,
\end{eqnarray}

\subsubsection{Quantum evolution in vicinity of degeneracy.}

To study the tunneling process near degeneracy, we assume $|u(t)\rangle$ be a solution of Eq. (\ref{Sch1e}) with the initial state at $t=0$ chosen as $|u_{\uparrow}\rangle$, and the final state of the system at a later time $t$ be $|u_{\uparrow}\rangle$ or $|u_{\downarrow}\rangle$. Then following \cite{BHC}, we compute the probability $P_{\uparrow\uparrow}$ $(P_{\downarrow\uparrow}) $ that the system is in the state  $|u_{\uparrow}\rangle$ ($|u_{\downarrow}\rangle$), respectively, as
\begin{eqnarray}\label{T3}
&P_{\uparrow\uparrow} =  |\cos(\Omega t/2)-i\cos\chi\,\sin(\Omega t/2)|^2 e^{-\lambda t} \\
&P_{\downarrow\uparrow} =  |\sin\chi\,\sin(\Omega t/2)|^2 e^{-\lambda t} \end{eqnarray}

In what follows we restrict ourselves by the case $\omega = \Delta $. Then, in according to the classification of the Tabl.1, the fictitious hyperbolic monopole emerges in the parameter space $\mathbb R^3 $ defined by the parameters of the system $\Re V_0$, $\Im V_0$, $\delta \in \mathbb R^3$.

There are two different regimes dependent on the relation between $\rho$ and $\delta$. For $\rho > \delta$ we have {\em one-sheeted hyperbolic monopole} and {\em coherent} tunneling process
\begin{eqnarray}\label{T4}
&P_{\uparrow\uparrow} =  e^{-\lambda t}\Big(\cos\frac{\Omega_0 t}{2}- \frac{\delta}{\Omega_0}\sin\frac{\Omega_0 t}{2}\Big)^2 , \\
&P_{\downarrow\uparrow} =   e^{-\lambda t} \frac{\rho^2}{\Omega_0^2}\sin^2\frac{\Omega_0 t}{2},
\end{eqnarray}
where $\Omega_0= |\rho^2 - \delta^2|^{1/2}$ denotes the Rabi frequency.

On the other hand, for $\rho < \delta$, there is  {\em incoherent} tunneling
\begin{eqnarray}\label{T5}
&P_{\uparrow\uparrow} =  e^{-\lambda t}\Big(\cosh\frac{\Omega_0 t}{2}- \frac{\delta}{\Omega_0}\sinh\frac{\Omega_0 t}{2}\Big)^2 , \\
&P_{\downarrow\uparrow} =   e^{-\lambda t} \frac{\rho^2}{\Omega_0^2}\sinh^2\frac{\Omega_0 t}{2}
\end{eqnarray}
and the associated monopole is the {\em two-sheeted hyperbolic monopole}. At the exceptional point we have $\Omega_0 = 0$, and both regimes yield
\begin{eqnarray}\label{T6}
P_{\uparrow\uparrow} =  \Big(1-  \frac{\delta t}{2}\Big)^2 e^{-\lambda t} \quad {\rm and} \quad P_{\downarrow\uparrow} = \Big(\frac{\delta t}{2}\Big)^2 e^{-\lambda t}.
\end{eqnarray}

\begin{figure}[tbh]
%\begin{minipage}[]{16cm}
%\begin{center}
%\psfrag{\rho}{\Huge $ \rho$}
%\psfrag{z}{\huge $ \mathbf z$}
\scalebox{0.325}{\includegraphics{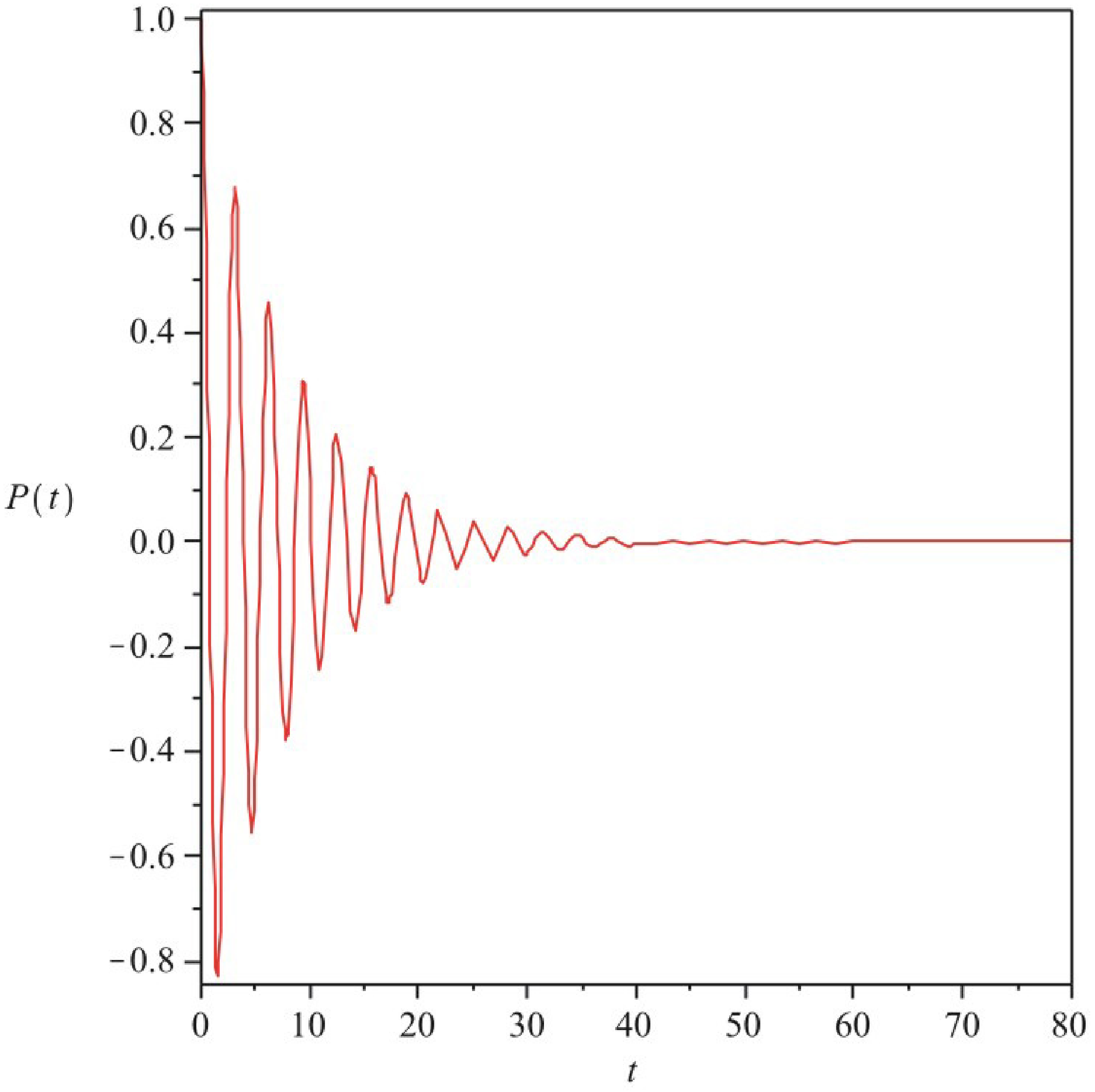}}
\hspace{1cm}
\scalebox{0.325}{\includegraphics{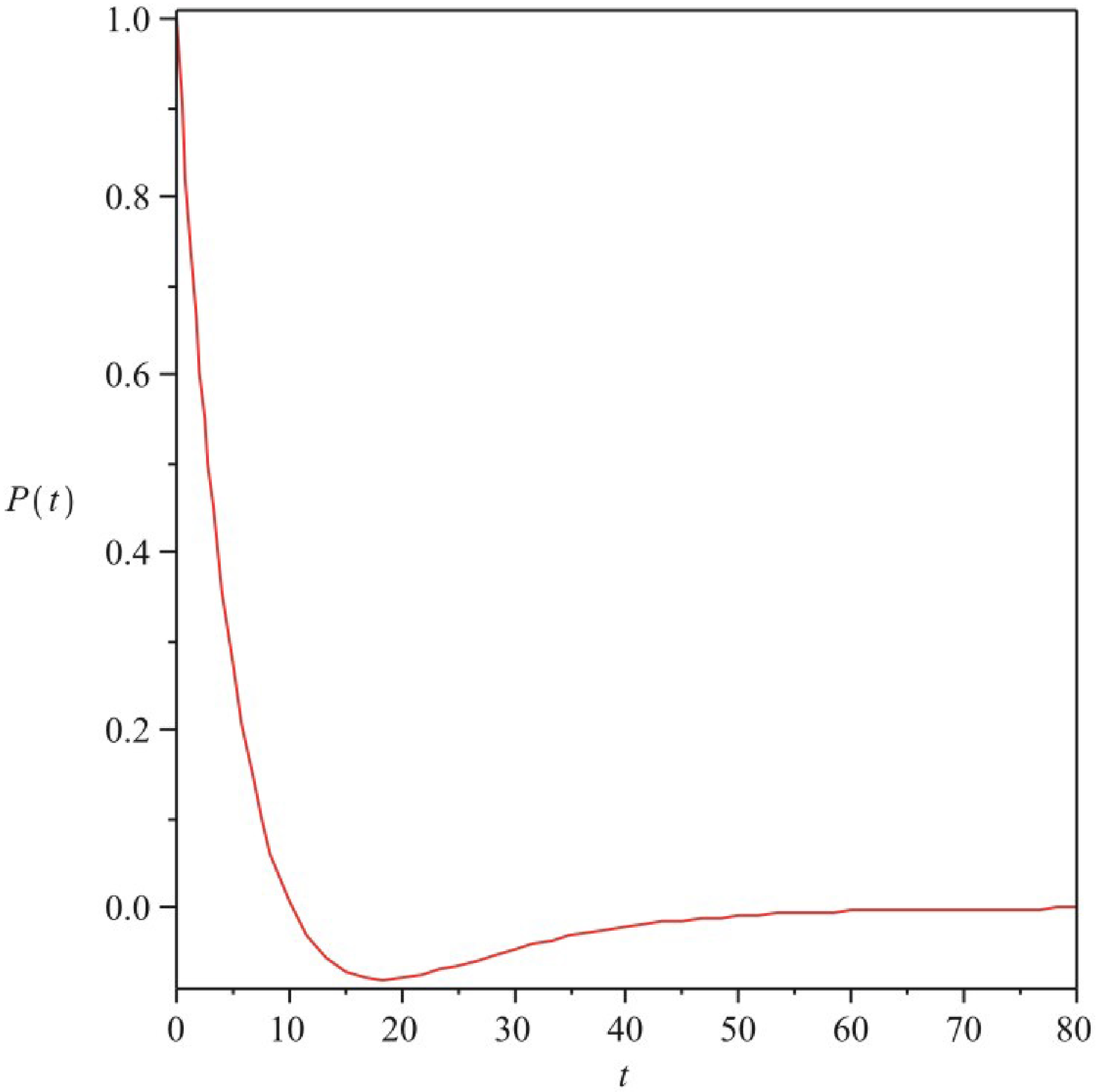}}
%\end{center}
\caption{Left panel: the Rabi oscillation $P(t)$ as function of $t$ for
coherent tunneling  ( $\rho > \delta$, $\Omega_0=0.025, \, \delta=0.1,\, \lambda = 0.125$). It is manifested as the quantum oscillation between the up and down states, $|u_{\uparrow}\rangle$ and $|u_{\downarrow}\rangle$. Right panel: $P(t)$ versus $t$ for
incoherent tunneling ( $\rho < \delta$, $\Omega_0=2, \, \delta=0.1,\, \lambda = 0.125$).
}\label{RO}
%\end{minipage}
\end{figure}

The Rabi oscillations is manifested as the quantum oscillation between up and down states and can be characterized by the following function:
$ P(t) = P_{\uparrow\uparrow} - P_{\downarrow\uparrow}$ \cite{WL}. The computation yields
\begin{eqnarray}\label{Eq20b}
&P(t) = e^{-\lambda t}\Big(\cos({\Omega_0 t})- \frac{\delta}{\Omega_0}\sin({\Omega_0 t})\Big), \; {\rm if} \; \rho > \delta \\
&P(t) = e^{-\lambda t}\Big(\cosh({\Omega_0 t}) - \frac{\delta}{\Omega_0}\sinh({\Omega_0 t})\Big), \; {\rm if} \; \rho < \delta
\end{eqnarray}
and at the exceptional point we obtain
\begin{eqnarray}\label{Eq20a}
P(t) = e^{-\lambda t}\Big(1- \frac{\delta t}{2}\Big).
\end{eqnarray}
The Rabi oscillation function $P(t)$ is plotted in Fig. \ref{RO}. In addition, it is simple to show that in the absence of dissipation $P(t) = \cos(\Omega_0 t)$.

Once return to the geometric phase defined by Eq.(\ref{GP8}), we obtain
\begin{eqnarray}
\label{GP17}
\fl
\gamma = \frac{\omega(\delta^2 + \Omega^2_0)}{2\Omega^3_0}(\sin \Omega_0 t - \Omega_0 t) -i\frac{\delta t}{2} + \frac{i}{2} \ln\frac{\Omega_0   + \delta\tan\frac{\Omega_0 t}{2}}{\Omega_0   - \delta\tan\frac{\Omega_0 t}{2}}, \quad {\rm if} \quad \rho > \delta,
\end{eqnarray}
and
\begin{eqnarray}
\label{GP18}
\fl
\gamma = \frac{\omega(\delta^2 + \Omega^2_0)}{2\Omega^3_0}(\sinh \Omega_0 t - \Omega_0 t) -i\frac{\delta t}{2} + \frac{i}{2} \ln\frac{\Omega_0   + \delta\tanh\frac{\Omega_0 t}{2}}{\Omega_0   - \delta\tanh\frac{\Omega_0 t}{2}}, \quad {\rm if} \quad \rho < \delta.
\end{eqnarray}

It follows from Eq.(\ref{GP17}) that the real part of the geometric phase $\Re\gamma(t)$ has the jump discontinuity $\Delta \Re\gamma =\mp \pi/2$ at the points
\begin{eqnarray}
t_n = \frac{2}{\Omega_0}\bigg( \pi n \pm \arctan\frac{\Omega_0}{\delta}\bigg), \; n=0,1, \dots
\end{eqnarray}
with pulse duration $\Delta t = t_{n+1} - t_n$ given by
\begin{eqnarray}
\Delta t = \frac{2\pi}{\Omega_0}\bigg( 1- \frac{2}{\pi} \arctan\frac{\Omega_0}{\delta}\bigg)
\end{eqnarray}
For the incoherent tunneling defined by Eq.(\ref{GP18}) the jump discontinuity $\Delta \Re\gamma =- \pi/2$ occurs at the point
\begin{eqnarray}
t_0 = \frac{2}{\Omega_0}\tanh^{-1}\bigg(\frac{\Omega_0}{\delta}\bigg)
\end{eqnarray}
\begin{figure}[tbh]
%\begin{minipage}[]{16cm}
%\begin{center}
%\psfrag{\rho}{\Huge $ \rho$}
%\psfrag{z}{\huge $ \mathbf z$}
\scalebox{0.325}{\includegraphics{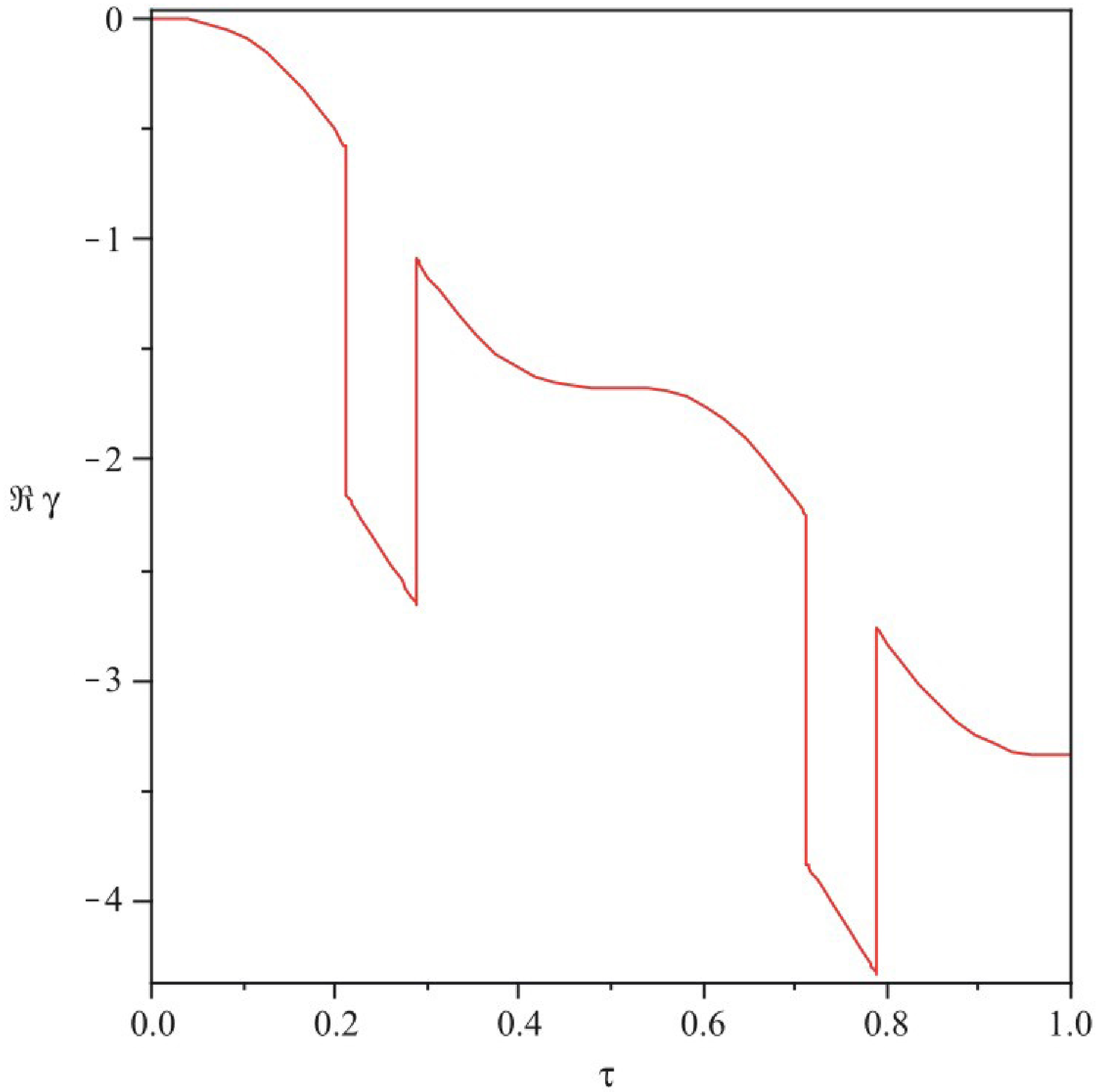}}
\hspace{1cm}
\scalebox{0.325}{\includegraphics{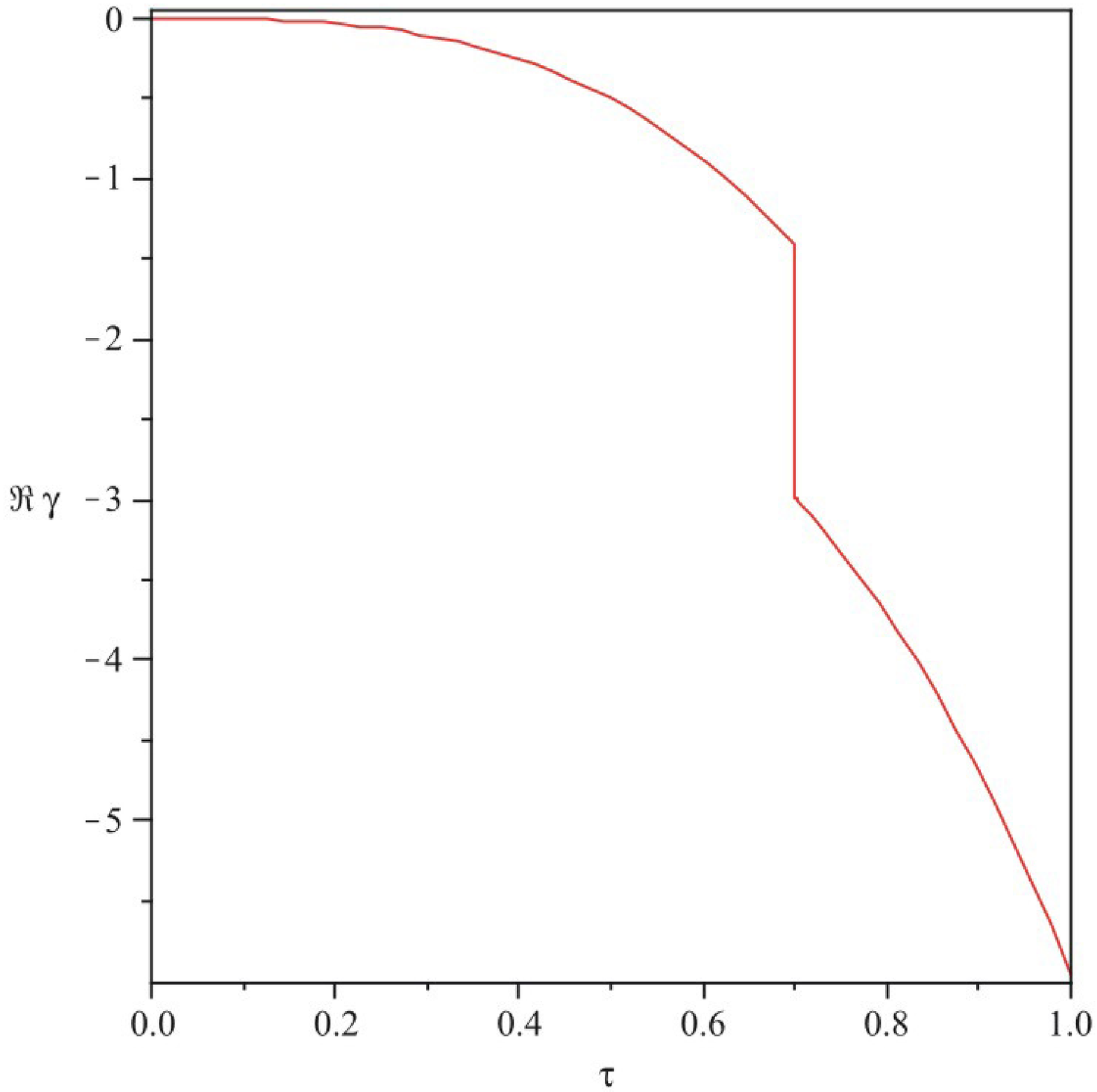}}
%\end{center}
\caption{ The real part of the geometric phase $\Re \gamma(\tau)$ versus time $\tau = 2\pi t/\omega$. Left panel (one-sheeted hyperbolic monopole): coherent tunneling $(\Omega_0=2, \, \delta=0.5,$ $ \omega = 1)$. Right panel (two-sheeted hyperbolic monopole): incoherent tunneling $(\Omega_0= 0.25,$ $ \delta=0.5,$ $\omega = 1)$.
}\label{RGEV}
%\end{minipage}
\end{figure}

\begin{figure}[tbh]
%\begin{minipage}[]{16cm}
\begin{center}
%\psfrag{Z}{\huge $\boldsymbol \Re \boldsymbol \gamma$}
%\psfrag{rho}{\huge $ \boldsymbol \rho$}
%\psfrag{tau}{\huge $ \boldsymbol \tau$}
\scalebox{0.375}{\includegraphics{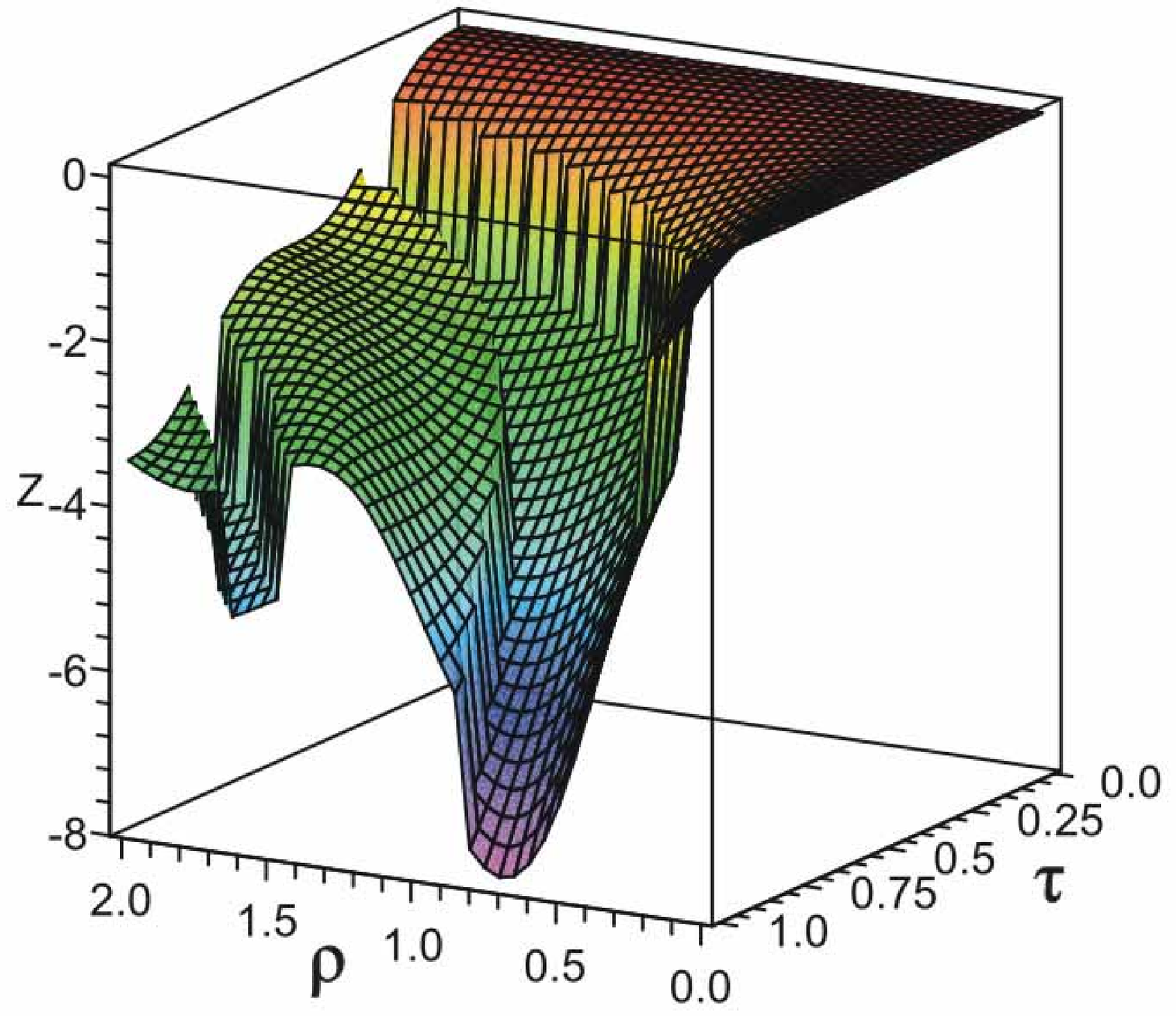}}
%\hspace{1cm}
\scalebox{0.375}{\includegraphics{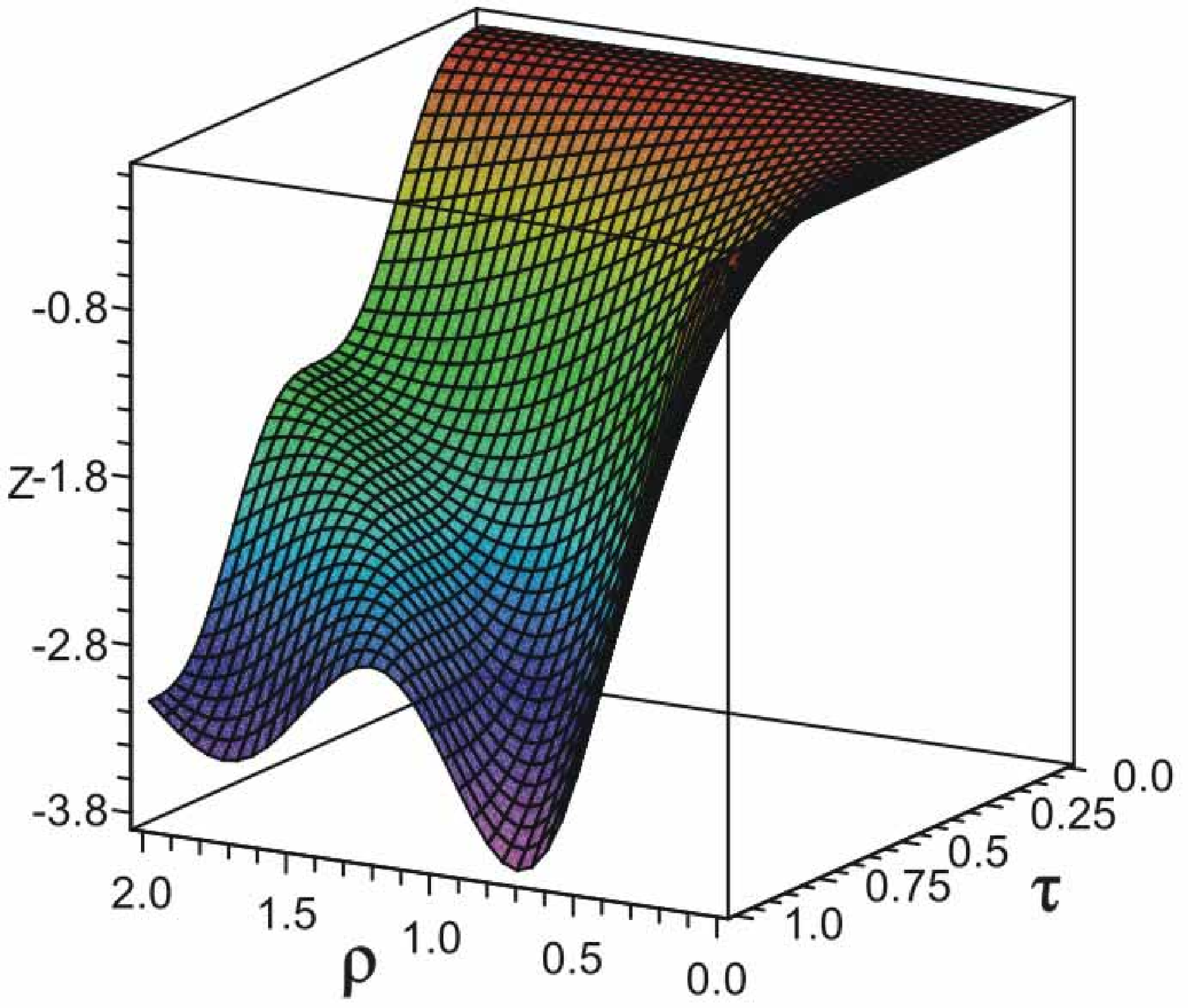}}
\end{center}
\caption{ Hyperbolic monopole. The real part of the geometric phase $\Re \gamma(\tau)$ versus time $\tau = 2\pi t/\omega$ and $\rho$ is depicted. Left panel: dissipative system $\delta \neq 0$ ( $\delta=0.5$, $ \omega = 1$). Right panel:  $\Re \gamma(\tau)$ is plotted in absence of dissipation ($\delta = 0$, $\omega = 1$). As can be seen from the plot, the pulses presented at the left are disappeared.
}\label{RGEV2}
%\end{minipage}
\end{figure}
At the exceptional point $\Omega_0=0$ we have $t_0= 2/\delta$ and the pulse duration $\Delta t \rightarrow \infty$. In absence of dissipation ( $\delta = 0$) the pulses are disappeared. Indeed, we have $t_0= 2/\delta \rightarrow \infty$ and the pulse duration $\Delta t \rightarrow 0$ while $\delta t \rightarrow 0$. The real part of the non-adiabatic geometric phase $\Re\gamma(\tau)$ as function of the time $\tau = 2\pi t/\omega$ is plotted in Fig. \ref{RGEV}, and in Fig.\ref{RGEV2} the real part of the geometric phase $\Re\gamma(\tau,\rho)$ versus $\tau$ and $\rho$ is depicted.

To conclude, we note that the transition emerged at the exceptional $\rho = \delta$ between two tunneling regimes is the topological phase transition in the parameter space, which can be described as follows: two-sheeted hyperbolic monopole $(\rho < \delta)$ $\leftrightarrow$ one-sheeted hyperbolic monopole  $(\rho > \delta)$.

\section{Conclusion}

In the present paper we considered the geometric phase and quantum tunneling in vicinity of diabolic and exceptional points. It has been shown the complex geometric phase associated with the degeneracy points is defined by the flux of complex fictitious `magnetic' monopole. In weak-coupling limit the leading contribution to the real part of geometric phase is given by the flux of the Dirac monopole plus quadrupole term, and the expansion for its imaginary part starts with the dipolelike field. Recently similar result has been obtained for a two-level spin-half system in a slowly varying magnetic field and weakly coupled to a dissipative environment \cite{WMSG}.

We found that the real part of  the complex geometric phase has a discontinuity jump at the exceptional point. We also have shown that the exceptional point is the critical point of the quantum-mechanical system, and at the latter the topological phase transition in the parameter space occurs.

Studying tunneling process near and at exceptional point we found two different regimes:  coherent and incoherent. The coherent tunneling is characterized by the Rabbi oscillations, also known as quantum echoes, and it has been shown that the dissipation brings into existence of pulses in the real part of the geometric phase. At exceptional point both tunneling regimes yield the quadratic dependence on time, that is in accordance with the results obtained in \cite{SBB,DFMR} for some specific non-Hermitian systems. The decay behavior predicted by Eqs. (\ref{T4}) - (\ref{T6}) has been recently observed in the experiment with a dissipative microwave billiard \cite{DFMR}.

Emerging of pulses in the geometric phase is a novel quantum phenomenon, which reflects the monopole structure of the system. This complex-valued phase effect may be detected by using the quantal dissipative interferometry \cite{SPEA,SE}. Note, that such a strong coupling effect of the environment should take place in generic dissipative systems, since in the neighborhood of the exceptional point only terms related to the invariant subspace formed by the two-dimensional Jordan block make substantial contributions and therefore the $N$-dimensional problem becomes effectively two-dimensional \cite{Arn,KMS}. We conclude by remarking that obtained results would be important in implementation of fault-tolerant quantum computation, and it is necessary to understand better the relation between geometric phase and decoherence to perform computation \cite{JVEC,ZW}.

\section*{Acknowledgements}

We thank A. B. Klimov, J. L. Romero and S. G. Ovchinnikov for helpful
discussions and comments. This work is supported by research grants SEP-PROMEP 103.5/04/1911 and CONACyT U45704-F.

\appendix
\section{Geometric phase for general evolution on the complex Bloch sphere}

In this Appendix we derive from the general definition of the geometric phase (\ref{GP5}) written for two-level system as
\begin{eqnarray}
\label{A1}
\gamma(\tau) = \frac{i}{2} \ln \left(\frac{\langle\tilde u(\tau)|u(0) \rangle}{\langle\tilde u(0)|u(t) \rangle}\right) + i \int_0^\tau \langle \tilde u(t)|\frac{d}{dt}|u(t) \rangle dt
\end{eqnarray}
the formula (\ref{Eq28b}) for computation of the geometric phase in terms of the complex Bloch vector. Eq. (\ref{A3}) generalizes to a non-hermitian Hamiltonian the formula obtained by Zang and Wang for computation of nonadiabatic noncyclic geometric phase for the Hermitian two-level system \cite{ZW5}.

{\bf Theorem}.  The complex geometric phase defined in Eq. (\ref{A1}) is given on the complex Bloch sphere $S^2_c$ by
\begin{eqnarray}\label{A2}
\fl \gamma (\tau) = -\frac{1}{2}\int_0^\tau \frac{n_1 \dot n_2 -n_2 \dot n_1}{1 + n_3}\, dt + \arctan\left(\frac{\sin(\beta_f - \beta_i)}{\cot(\alpha_f/2)\cot(\alpha_i/2) + \cos(\beta_f - \beta_i)}\right),
\end{eqnarray}
where integration is performed along the unique curve $\mathbf n(t)$ on the unit sphere $S^2_c$, joining the initial point $\mathbf n(0) = (\sin \alpha_i\cos\beta_i,\sin \alpha_i\sin\beta_i, \cos\alpha_i)$ and the final point $\mathbf n(\tau) = (\sin \alpha_f\cos\beta_f,\sin \alpha_f\sin\beta_f, \cos\alpha_f)$.\\

\noindent
{\em Proof.} In general form, for a two-level system in terms of column and row vectors we have
\begin{eqnarray}\label{A3}
|u(t) \rangle = \left(\begin{array}{c}
                  a(t) \\
                  b(t)
                \end{array}\right), \quad \langle \tilde u(t)| = (\tilde a(t),\tilde b(t)).
\end{eqnarray}
After some algebra and using the definition of the Bloch vector $\mathbf n(t)= \langle\widetilde\Psi(t)|\boldsymbol\sigma|\Psi(t)\rangle$, we find
\begin{eqnarray}\label{B3}
n_1(t) = a\tilde b+ \tilde a b, \quad
n_2(t) = i(a\tilde b - \tilde a b), \quad
n_3(t) = a\tilde a - b\tilde b
\end{eqnarray}
From here, setting $\mathbf n(t) = (\sin \alpha\cos\beta,\sin \alpha\sin\beta, \cos\alpha)$, we obtain
\begin{eqnarray}\label{A4}
a\tilde b = \sin\frac{\alpha}{2} \cos\frac{\alpha}{2} e^{-i\beta},& \quad a\tilde a = \cos^2\frac{\alpha}{2},  \\
\tilde a b = \sin\frac{\alpha}{2} \cos\frac{\alpha}{2} e^{i\beta},&
\quad b\tilde b = \sin^2 \frac{\alpha}{2}.
\label{A5}
\end{eqnarray}

Next, denoting by $|u_i \rangle$ and $|u_f \rangle$ initial and final states, respectively, we can write the total phase as follows:
\begin{eqnarray}
\label{A6}
\gamma_t =\frac{i}{2} \ln \left(\frac{\langle\tilde u(\tau)|u(0) \rangle}{\langle\tilde u(0)|u(t) \rangle}\right)= \frac{i}{2} \ln\left(\frac{\langle\tilde u_f|u_i \rangle}{\langle\tilde u_i|u_f \rangle}\right) =  \frac{i}{2} \ln\left(\frac{\tilde a_f a_i + \tilde b_f b_i}{\tilde a_i a_f + \tilde b_i b_f}\right)
\end{eqnarray}
Then, applying (\ref{A4}) and (\ref{A5}), we obtain
\begin{eqnarray}
\label{A7}
\gamma_t = \frac{i}{2} \ln\left(\frac{\cot(\alpha_f/2)\cot(\alpha_i/2) + e^{i(\beta_i - \beta_f)}}{\cot(\alpha_f/2)\cot(\alpha_i/2) + e^{i(\beta_f - \beta_i)}}\right) +  \frac{i}{2} \ln\left(\frac{a_i\tilde a_f}{a_f\tilde a_i}\right).
\end{eqnarray}
This yields
\begin{eqnarray}
\label{A8}
\fl
\gamma_t = \arctan\left(\frac{\sin(\beta_f - \beta_i)}{\cot(\alpha_f/2)\cot(\alpha_i/2) + \cos(\beta_f - \beta_i)}\right) + \frac{i}{2} \int_0^\tau \left(\frac{d\tilde a}{\tilde a}- \frac{d a}{a}\right).
\end{eqnarray}
Since the dynamical phase
\begin{eqnarray}
\label{A9}
\gamma_d = - i \int_0^\tau \langle \tilde u(t)|\frac{d}{dt}|u(t) \rangle dt = -i\int_0^\tau (\tilde a \dot a + \tilde b \dot b) dt,
\end{eqnarray}
we obtain
\begin{eqnarray}
\label{A10}
\gamma =\gamma_t- \gamma_d =& \arctan\left(\frac{\sin(\beta_f - \beta_i)}{\cot(\alpha_f/2)\cot(\alpha_i/2) + \cos(\beta_f - \beta_i)}\right) \nonumber \\
&+  i\int_0^\tau \left((\tilde a \dot a + \tilde b \dot b)+ \frac{1}{2}\left(\frac{\dot {\tilde a}}{\tilde a}- \frac{\dot a}{a}\right)\right)dt.
\end{eqnarray}
Using the relations Eqs. {(\ref{B3}) -- (\ref{A5})}, we find
 \begin{eqnarray}
\label{A11}
(1- \cos\alpha)\dot\beta= \frac{n_1 \dot n_2 -n_2 \dot n_1}{1 + n_3}= -2i \left((\tilde a \dot a + \tilde b \dot b)+ \frac{1}{2}\left(\frac{\dot {\tilde a}}{\tilde a}- \frac{\dot a}{a}\right)\right).
\end{eqnarray}
Then, inserting this result into (\ref{A10}), we obtain
\begin{eqnarray}\label{A12}
\fl \gamma(\tau) = -\frac{1}{2}\int_0^\tau \frac{n_1 \dot n_2 -n_2 \dot n_1}{1 + n_3}\, dt + \arctan\left(\frac{\sin(\beta_f - \beta_i)}{\cot(\alpha_f/2)\cot(\alpha_i/2) + \cos(\beta_f - \beta_i)}\right).
\end{eqnarray}

$\hspace{14cm}\Box$
~~~~~~~~\\
{\bf Corollary}. The geometric phase can be calculated by the following integral
\begin{eqnarray}\label{A13}
\fl \gamma(\tau) = -\frac{1}{2}\int_0^\tau (1- \cos\alpha)\dot\beta dt + \arctan\left(\frac{\sin(\beta_f - \beta_i)}{\cot(\alpha_f/2)\cot(\alpha_i/2) + \cos(\beta_f - \beta_i)}\right).
\end{eqnarray}

\section*{References}

\bibliography{dep_arxiv}
%\bibliography{EP,nonass}
%\bibliography{nonass}

\end{document}